\documentclass[11pt]{iopart}
\usepackage[T1]{fontenc}
\usepackage[latin1]{inputenc}
\usepackage{graphicx}

\makeatletter

\providecommand{\tabularnewline}{\\}

\newenvironment{lyxlist}[1]
{\begin{list}{}
{\settowidth{\labelwidth}{#1}
 \setlength{\leftmargin}{\labelwidth}
 \addtolength{\leftmargin}{\labelsep}
 }}
{\end{list}}

\usepackage{psfrag}
\usepackage[english]{babel}
\makeatother

\begin{document}

\title{On the range of validity of the fluctuation theorem for stochastic Markovian dynamics}

\author{A R\'{a}kos$^{1,2}$ and R J Harris$^{3,4}$}

\address{$^1$ Department of Physics of Complex Systems, Weizmann Institute of Science,
Rehovot 76100, Israel}
\address{$^2$ Research Group for Condensed Matter Physics of the Hungarian Academy of Sciences, Budafoki u.\ 8, 1111 Budapest, Hungary}

\address{$^3$ Fachrichtung Theoretische Physik, Universit\"at des Saarlandes, 66041 Saarbr\"ucken, Germany}
\address{$^4$ Present Address: School of Mathematical Sciences, Queen Mary, University of London, Mile End Road, London, E1 4NS, United Kingdom}
\eads{\mailto{rakos@phy.bme.hu}, \mailto{rosemary.harris@qmul.ac.uk}}

\begin{abstract}
We consider the fluctuations of generalized currents in stochastic Markovian dynamics. The large deviations of current fluctuations are shown to obey a Gallavotti-Cohen (GC) type symmetry in systems with a finite state space. However, this symmetry is not guaranteed to hold in systems with an infinite state space. A simple example of such a case is the Zero-Range Process (ZRP). Here we discuss in more detail the already reported \cite{Harris2006} breakdown of the GC symmetry in the context of the ZRP with open boundaries and we give a physical interpretation of the phases that appear.
Furthermore, the earlier analytical results for the single-site case are extended to cover multiple-site systems.
We also use our exact results to test an efficient numerical algorithm of Giardin\`a, Kurchan and Peliti~\cite{Giardina06}, which was developed to measure the current large deviation function directly. We find that this method breaks down in some phases which we associate with the gapless spectrum of an effective Hamiltonian.
\end{abstract}
\tableofcontents
\markboth{\hfill}{\hfill}

\section{Introduction}

An important step in the understanding of non-equilibrium systems has
been the development of so-called fluctuation
theorems~\cite{Evans02b,Ritort03} which quantify irreversibility by
relating the probability of some ``backward'' process to that of a
corresponding ``forward'' process.  Formally, these fluctuation
relationships all derive from a statement about the relative
probabilities of a trajectory in phase space and the time-reversed
trajectory.  The various theorems found in the literature can be
divided into two broad classes: transient relations (which are exact
for finite times) and steady-state relations (which hold only
asymptotically in the long-time limit)---see,
e.g.,~\cite{Kurchan07,Harris2007a} for recent discussion of the
various interconnections.  In the present paper we will mainly be
interested in asymptotic symmetries, specifically whether (or not) a
given quantity satisfies a relation of the form
\begin{equation} 
\frac{\mathcal{P}(-r,t)}{\mathcal{P}(r,t)} \sim \rme^{-r t}
\label{e:GCFT}.
\end{equation} 
Here $\mathcal{P}(r,t)$ is the probability to observe, over time interval $[0,t]$, a time-averaged value
$r\equiv R/t$ of some time-extensive quantity $R$ (e.g., entropy production).  The symbol $\sim$ means logarithmic equality in the limit of large
$t$.  In this paper, we refer to the relationship~(\ref{e:GCFT}) as ``the
Gallavotti-Cohen fluctuation theorem'' or simply ``the fluctuation theorem''.

Historically, the concept of a fluctuation theorem emerged from
computer simulations investigating the entropy production fluctuations
in a sheared fluid~\cite{Evans93}.  A rigorous derivation of the
form~(\ref{e:GCFT}) was then given for the steady state of
deterministic systems~\cite{Gallavotti95,Gallavotti95b} where the
entropy-like functional $r$ can be identified with the rate of phase
space contraction.  Later proofs addressed stochastic
dynamics~\cite{Kurchan98, Lebowitz99}, leading to a general property
of the large deviation function sometimes referred to as
``Gallavotti-Cohen symmetry'' (or GC symmetry for short).  Here
$r$ is associated with a current of some quantity through the
system (e.g., the particle current in a lattice gas) and, for bounded
state-space, the relation~(\ref{e:GCFT}) holds for arbitrary initial
condition.  Recent work by Kurchan~\cite{Kurchan2007} has also explored how to recast the deterministic fluctuation theorem as the vanishing-noise limit of the stochastic one.  In parallel to the theoretical development there has been
much successful work on experimental verification of fluctuation
theorems; for reviews see~\cite{Ritort03,Narayan04,Zamponi07}. In
particular, we note that recent experiments using an optical defect in
diamond provide a simple realization of a two-state stochastic
system~\cite{Schuler05,Tietz06b}.

Very recently there has been considerable theoretical, experimental
and numerical interest in cases where the symmetry~(\ref{e:GCFT})
breaks down, see
e.g.,~\cite{VanZon03,Garnier05,Dolowschiak05,Bonetto06b,Harris2006,Puglisi06,Visco06}.
It is now understood that this effect can be attributed to certain
boundary terms which become relevant in the case of infinite state
space; see section~\ref{s:GC} below
for a detailed exposition in the context of stochastic Markovian
dynamics.  For deterministic systems, the effect of unbounded
potentials was discussed by Bonetto~\emph{et al.}~\cite{Bonetto06b}.
They argued for the restoration of the symmetry by removal of the
``unphysical'' singular terms.  Earlier a similar phenomenon was found
for a model of a trapped-Brownian particle treated via a Langevin equation~\cite{VanZon03,VanZon04}.
Within the Langevin framework, discussion of some related subtleties can also be found
in~\cite{Farago02,Baiesi06}.

In~\cite{Harris2006} we provided a general discussion of this breakdown of
Gallavotti-Cohen symmetry for \emph{many-particle} stochastic dynamics
within the context of a particular model---the zero-range process
(ZRP).  The ZRP plays a paradigmatic role in non-equilibrium
statistical mechanics, hence such work offers a contribution to
general understanding as well as providing concrete results for an
important model.  In the present paper, we give details of the
analytical calculations leading to those results 
and compare them with direct evaluation of the current large deviation
function using the recent algorithm of Giardin\`a, Kurchan and
Peliti~\cite{Giardina06}.  We also discuss some 
generalizations and present new results for the multi-site ZRP.

Interestingly, in the zero-range process, we do not find a constant
value for the ratio of probabilities for large forward and backward
current fluctuations.  This is in stark contrast to analytical
arguments~\cite{VanZon03,Bonetto06b,Baiesi06} and numerical
work~\cite{Puglisi06} for other models.  We argue below that the
failure to observe this form of ``extended fluctuation
theorem'' is due to strong correlations in our model between the boundary terms and
the integrated current.  These correlations persist even in the
long-time limit; it would be interesting to see if effects of this
type can be observed in any experimental situations.

The plan of the rest of the paper is as follows.
First, in section~\ref{s:GC}, we give a
general derivation of the Gallavotti-Cohen fluctuation theorem for stochastic
systems and indicate also the relation to transient fluctuation
theorems.  Then, in section~\ref{s:ZRP}, we introduce our zero-range
model and outline its treatment within this general framework.
section~\ref{s:calc} is devoted to a detailed calculation of the
current large deviation function for the single-site version of this
model, giving a concrete example of the breakdown of GC symmetry.  In
section~\ref{s:num} the analytical approach is complemented by
use of the algorithm of~\cite{Giardina06} to obtain new
numerical results for the large deviation function (and general insight into the applicability of the ``cloning'' method).  Significantly,
in section~\ref{s:large} we extend our discussion to larger systems,
demonstrating how information about the current fluctuations in an
$L$-site ZRP can be obtained by mapping to an effective single-site
model.  Section~\ref{s:conc} contains some conclusions and general
perspectives.

\section{The fluctuation theorem for Markovian dynamics}

\label{s:GC}

\subsection{Central argument}

Here we present a derivation of the fluctuation theorem for generalized
currents of Markov processes defined on a finite configuration space.
Our argument is based on that of \cite{Lebowitz99}. 

Consider a continuous time Markov process which satisfies detailed
balance in the stationary state. The system can be described by the
transition rates $w_{\sigma'\sigma}^{{\normalcolor {\normalcolor {\normalcolor \mathrm{eq}}}}}$
from configuration $\sigma$ to $\sigma'$. In addition, consider
a counter $J$, the value of which increases by the amount $\Theta_{\sigma'\sigma}$
at each transition $\sigma\to\sigma'$. Here the matrix $\Theta$,
which is required to be real and antisymmetric, can describe any type
of real or abstract current in the system. As an example one can consider
the particle current through a given bond: in this case $\Theta_{\sigma'\sigma}$
is the number of particles hopping across the bond at a transition
$\sigma\to\sigma'$ (which can be positive or negative depending on
the direction of hopping). At $t=0$ the counter $J$ is set to zero;
for any positive time it is a functional acting on the paths (sequences
of configurations) from time $0$ to time $t$ and can be written
as 
\begin{equation}
J(t,\{\sigma\})=\sum_{k=1}^{n-1}\Theta_{\sigma_{k+1}\sigma_{k}}.\label{eq:action}
\end{equation}
Here $n$ is the number of configurations $\sigma_k$ visited during time $t$.
In the following we refer to $J$ as the ``time-integrated current'' (or, where no confusion can arise, simply  as the ``current''). 
Note that, due to detailed balance, the mean of this current $\langle J(t)\rangle$
is always zero in equilibrium.

We define the driven system by the modified transition rates 
\begin{equation}
w_{\sigma'\sigma}=w_{\sigma'\sigma}^{\mathrm{eq}}\rme^{\frac{E}{2}\Theta_{\sigma'\sigma}}\label{eq:wdriven}
\end{equation}
from configuration $\sigma$ to $\sigma'$, where $E$ is a driving
field conjugated to the specific current under consideration. In what
follows we show that in this driven system the probability distribution
function $\mathcal{P}(J,t)$ of the random variable $J(t)$ satisfies the
relation 
\begin{equation}
\frac{\mathcal{P}(J,t)}{\mathcal{P}(-J,t)}\sim \rme^{EJ}\label{eq:FR}
\end{equation}
asymptotically for large times, provided the state space (i.e., the
number of possible configurations) is finite. Here the power $EJ$
can be interpreted as the work done on the system by the external
field. This is the statement of the fluctuation theorem. 

In cases where the sum $\sum_n \Theta_{\sigma_{n}, \sigma_{n+1}}$ gives zero for all periodic paths $\sigma_n$ in the configuration space, $J$ becomes a simple function of the initial and final configuration, i.e., there is no real dependence on the history. In this special case not only the original but also the above-defined ``driven'' system would satisfy detailed balance in the stationary state.  In the following we assume that this is not the case, i.e., there are periodic paths in the configuration space for which the sum $\sum_{n} \Theta_{\sigma_{n}, \sigma_{n+1}}$ is non-zero.

For a given \emph{non-equilibrium} model with rates $w_{\sigma'\sigma}$
one can apply a reversed argument. In this case physically one can
think of $E$ as a negative driving field, conjugated to the current
under consideration, which is needed in order to {}``restore'' detailed
balance. If, for a specific value of $E$, the rates $w_{\sigma'\sigma}^{\mathrm{eq}}$
(defined by (\ref{eq:wdriven})) lead to detailed balance then the
fluctuation relation (\ref{eq:FR}) holds for this current. This gives
some freedom for the quantity $J$ which enters the fluctuation relation.
The action functional of \cite{Lebowitz99} ($W$ in that paper)
corresponds to the specific choice $\Theta_{\sigma'\sigma}=\ln\frac{w_{\sigma'\sigma}}{w_{\sigma\sigma'}}$
with $E=1$, which leads to $w_{\sigma'\sigma}^{\mathrm{eq}}=\sqrt{w_{\sigma'\sigma}w_{\sigma\sigma'}}$.
These rates indeed lead to detailed balance, since for each pair of
configurations the forward and backward transition rates are equal.
This also implies that in the corresponding equilibrium distribution
every configuration has the same weight.

Let us now define the rate $w(\{\sigma\})$ for a full path as
\begin{equation}
w(\{\sigma\})=w_{\sigma_{1}\sigma_{0}}w_{\sigma_{2}\sigma_{1}}\cdots w_{\sigma_{n}\sigma_{n-1}}\rme^{-\frac{t_{0}}{\tau_{0}}}\rme^{-\frac{t_{1}-t_{0}}{\tau_{1}}}\cdots \rme^{-\frac{t-t_{n-1}}{\tau_{n}}},\label{eq:pathrate}
\end{equation}
 where 
$t_{k}$ denotes the time when the transition from configuration $\sigma_{k}$ to $\sigma_{k+1}$ happened
and $\tau_{k}=\left(\sum_{\sigma'}w_{\sigma'\sigma_{k}}\right)^{-1}$
corresponds to the overall exit rate from configuration $\sigma_k$. 
The conditional
probability of such a path with transition times between $t_{k}$
and $t_{k}+dt$, provided at time 0 the system starts in configuration
$\sigma_{0}$ is $w(\{\sigma\})dt^{n-1}$. Using (\ref{eq:wdriven})
one can readily show that 
\begin{equation}
\frac{p_{\sigma_{0}}^{\mathrm{eq}}w(\{\sigma\})}{p_{\sigma_{n}}^{\mathrm{eq}}w\left(\{\sigma\}^{\mathrm{rev}}\right)}=\rme^{EJ(t,\{\sigma\})},
\end{equation}
where $\{\sigma\}^{\mathrm{rev}}$ is the time-reversed path of $\{\sigma\}$
and $p_{\sigma}^{\mathrm{eq}}$ is the equilibrium probability of
configuration $\sigma$. In reference \cite{Taniguchi2007} the
above relation is referred to as the ``non-equilibrium detailed
balance condition'' with $EJ$ being the work done on the system.
This also suggests the identification of $EJ$ as the work.

We note that in the case of discrete time dynamics the above scenario
is very similar, the only difference is that here the quantities
$w_{\sigma'\sigma}$ and
$w(\{\sigma\})$ denote transition probabilities rather than rates
and the exponential factors in (\ref{eq:pathrate}) are replaced by
diagonal elements of the transition matrix $w$. 

\subsection{Proof of the asymptotic fluctuation theorem for finite systems}

In the calculation we use the so-called quantum Hamiltonian formalism where a basis vector is associated with each possible configuration and the state of the system (a probability measure on the configuration space) is denoted by a vector $|P\rangle$ in this space with the normalization $\langle s|P\rangle =1$. Here $\langle s|$ is a row vector with components $(1,1,1,\ldots)$. In this formalism the master equation takes the form 
\begin{equation}
\frac{\partial}{\partial t}|P\rangle = - H |P\rangle,
\end{equation}
which is similar to the Schr\"odinger equation in imaginary time. The transition rates are in the off-diagonal elements of $H$ and the conservation of probability requires 
\begin{equation}
\label{Hnormalization}
\langle s|H=0.
\end{equation}
For more details on this formalism see \cite{Schutz2001}.

As a first step of the proof we introduce the joint probability distribution function $\mathcal{P}_{\sigma}(J,t)$,
which denotes the probability of being in configuration $\sigma$
and having the value $J$ of the current at time $t$. In what follows
we calculate the generating function $\langle \rme^{-\lambda J(t)}\rangle.$
\begin{equation}
\left\langle \rme^{-\lambda J(t)}\right\rangle =\left\langle s|g(t)\right\rangle ,
\end{equation}
 where 
\begin{equation}
g(t)_{\sigma}=\sum_{J}\mathcal{P}_{\sigma}(J,t)\rme^{-\lambda J}.
\end{equation}
 The time derivative of $g(t)$ is 
\begin{eqnarray}
\frac{\rmd}{\rmd t}g(t)_{\sigma} & =\sum_{J}\sum_{\sigma'}\left(\mathcal{P}_{\sigma'}(J-\Theta_{\sigma\sigma'},t)w_{\sigma\sigma'}-\mathcal{P}_{\sigma}(J,t)w_{\sigma'\sigma}\right)\rme^{-\lambda J}\cr
 & =\sum_{J'}\sum_{\sigma'}\mathcal{P}_{\sigma'}(J',t)w_{\sigma\sigma'}\rme^{-\lambda J'}\rme^{-\lambda \Theta_{\sigma\sigma'}}-g(t)_{\sigma}\sum_{\sigma'}w_{\sigma'\sigma}\cr
 & =\sum_{\sigma'}g(t)_{\sigma'}w_{\sigma\sigma'}\rme^{-\lambda \Theta_{\sigma\sigma'}}-g(t)_{\sigma}\sum_{\sigma'}w_{\sigma'\sigma}\cr
 & =-\tilde{H}(\lambda)_{\sigma\sigma'}g(t)_{\sigma'},
\end{eqnarray}
where $\tilde{H}(\lambda)$ is a modified Hamiltonian in which the
transition rates corresponding to $\sigma'\to\sigma$ are multiplied
by the factor $\rme^{-\lambda \Theta_{\sigma\sigma'}}$. Note that $\tilde{H}(\lambda)$
is a non-stochastic matrix with $\langle s|\tilde{H}(\lambda)\neq\langle s|$
for $\lambda\neq0$. Since $|g(0)\rangle$ is identical to the initial
measure $|P_{0}\rangle$ the generating function takes the form
\begin{equation}
\left\langle \rme^{-\lambda J(t)}\right\rangle =\left\langle s\left|\rme^{-\tilde{H}(\lambda)t}\right|P_{0}\right\rangle .\label{eq:Htilde}
\end{equation}
The long-time behaviour of this quantity is characterized by the function
\begin{equation}
e(\lambda)=-\lim_{t\to\infty}\frac{1}{t}\ln\left\langle \rme^{-\lambda J(t)}\right\rangle .\label{eq:elambda}
\end{equation}
One finds from (\ref{eq:Htilde}) that as long as the configuration
space is finite, $e(\lambda)$ is given by the lowest eigenvalue $\epsilon_{0}(\lambda)$
of $\tilde{H}(\lambda)$, since 
\begin{equation}
\left\langle \rme^{-\lambda J(t)}\right\rangle =\sum_{i}\left\langle s|\psi_{i}\right\rangle \left\langle \psi_{i}|P_{0}\right\rangle \rme^{-\epsilon_{i}(\lambda)t}\simeq\left\langle s|\psi_{0}\right\rangle \left\langle \psi_{0}|P_{0}\right\rangle \rme^{-\epsilon_{0}(\lambda)t},\label{eq:as}
\end{equation}
Where $\epsilon_{i}(\lambda)$ are the eigenvalues and $\psi_{i}$
are the eigenvectors of $\tilde{H}(\lambda)$. It is easy to show
that $e(\lambda)$ is related to the large deviation function 
\begin{equation}
\hat{e}(j)=-\lim_{t\to\infty}\frac{1}{t}\ln \mathcal{P}(tj,t)
\end{equation}
of the time-averaged current $j=J/t$ through a Legendre transformation:
\begin{equation}
\hat{e}(j)=\max_{\lambda}\{ e(\lambda)-\lambda j\},\qquad e(\lambda)=\min_{j}\{\hat{e}(j)+\lambda j\}.\label{eq:Legendre}
\end{equation}

As a second step in the proof we show that $\tilde{H}$ has the symmetry
property 
\begin{equation}
\tilde{H}(\lambda)^{T}=P_{\mathrm{eq}}^{-1}\tilde{H}(E-\lambda)P_{\mathrm{eq}},\label{eq:Htildesymm}
\end{equation}
 where $P_{\mathrm{eq}}$ is a matrix with the equilibrium probabilities
$p_{\sigma}^{\mathrm{eq}}$ on the diagonal and zero elsewhere. For
the rhs of (\ref{eq:Htildesymm}) one finds
\begin{equation} \fl
\left[P_{\mathrm{eq}}^{-1}\tilde{H}(E-\lambda)P_{\mathrm{eq}}\right]_{\sigma'\sigma}=-w_{\sigma'\sigma}\rme^{-(E-\lambda)\Theta_{\sigma'\sigma}}\frac{p_{\sigma}^{\mathrm{eq}}}{p_{\sigma'}^{\mathrm{eq}}}(1-\delta_{\sigma'\sigma})+\sum_{\rho}w_{\rho\sigma}\delta_{\sigma'\sigma},
\end{equation}
which leads to 
\begin{equation}
w_{\sigma\sigma'}\rme^{-\lambda \Theta_{\sigma\sigma'}}=w_{\sigma'\sigma}\rme^{-(E-\lambda)\Theta_{\sigma'\sigma}}\frac{p_{\sigma}^{\mathrm{eq}}}{p_{\sigma'}^{\mathrm{eq}}}
\end{equation}
for the non-diagonal elements of equality (\ref{eq:Htildesymm}).
Using (\ref{eq:wdriven}) and the fact that the matrix $\Theta$ is antisymmetric,
the above condition takes the form 
\begin{equation}
w_{\sigma\sigma'}^{\mathrm{eq}}p_{\sigma'}^{\mathrm{eq}}=w_{\sigma'\sigma}^{\mathrm{eq}}p_{\sigma}^{\mathrm{eq}},
\end{equation}
which is just the detailed balance condition for the equilibrium system
and is trivially satisfied. This proves the relation (\ref{eq:Htildesymm})
(The diagonal part of both the lhs and rhs is just the diagonal part
of the original non-equilibrium Hamiltonian.) 

As a corollary of the symmetry property (\ref{eq:Htildesymm}) one
finds the important relation 
\begin{equation}
e(\lambda)=e(E-\lambda).\label{eq:esymmetry}
\end{equation}
For the large deviation function $\hat{e}(j)$, using the Legendre
transformation (\ref{eq:Legendre}) this implies 
\begin{equation}
\hat{e}(-j)=Ej+\hat{e}(j),
\end{equation}
which is equivalent to (\ref{eq:FR}).

\subsection{Transient fluctuation theorem}

The relation (\ref{eq:FR}) is true only asymptotically, for large
times. However, due to the symmetry (\ref{eq:Htildesymm}), one can show
that for specific initial conditions it can be made exact for any
finite time. Namely, taking the equilibrium distribution $|p^{\mathrm{eq}}\rangle$
as the initial condition then for the generating function one finds
\begin{eqnarray} \fl
\langle \rme^{-\lambda J(t)}\rangle=\langle s|\rme^{-\tilde{H}(\lambda)t}|p^{\mathrm{eq}}\rangle=\langle p^{\mathrm{eq}}|\rme^{-\tilde{H}(\lambda)^{T}}|s\rangle =
\langle p^{\mathrm{eq}}|P_{\mathrm{eq}}^{-1}\rme^{-\tilde{H}(E-\lambda)t}P_{\mathrm{eq}}|s\rangle
\cr =\langle s|\rme^{-\tilde{H}(E-\lambda)t}|p^{\mathrm{eq}}\rangle=\langle \rme^{-(E-\lambda)J(t)}\rangle.
\end{eqnarray}
This implies that the relation (\ref{eq:FR}) holds exactly for any
finite time (but only for this specific initial condition). This is
the statement of the transient fluctuation theorem of Evans and Searles
\cite{Evans1994, Searles1999, Evans02b}.

\subsection{Boundary terms}

As mentioned above, in a non-equilibrium system one has some
freedom to choose the current to be considered. Here we investigate
whether two different choices really give two independent relations
for the fluctuations of the non-equilibrium system. In order to satisfy
the conditions of the theorem, by applying $-E$ field in the non-equilibrium
model one should get back to detailed balance, which can be formulated
as 
\begin{equation}
\frac{w_{\sigma\sigma'}\rme^{-\frac{E}{2}\Theta_{\sigma\sigma'}}}{w_{\sigma'\sigma}\rme^{-\frac{E}{2}\Theta_{\sigma'\sigma}}}=\frac{\rme^{-V_{\sigma}}}{\rme^{-V_{\sigma'}}}.\label{eq:db12}
\end{equation}
 Here $V_{\sigma}$ is the energy of configuration $\sigma$
in the equilibrium system. Taking the logarithm of (\ref{eq:db12})
and summing up for a path results in 
\begin{equation}
J^{*}(t)-EJ(t)=V_{\sigma_{\mathrm{ini}}}-V_{\sigma_{\mathrm{fin}}},
\end{equation}
where $J^{*}$ is the action functional of \cite{Lebowitz99} and
$V_{\sigma_{\mathrm{ini(fin)}}}$ is the potential in the initial
(final) state. This shows that any two currents (that satisfy the
conditions of the fluctuation theorem) differ only in boundary terms.
In finite systems these terms are bounded, consequently their contribution
to the net current is negligible in the limit, where $t\to\infty$.
In these systems the fluctuation theorems for different currents give
essentially the same information. 

In order to demonstrate this, consider a driven exclusion process on
a finite lattice. Here the number of possible configurations is clearly
finite. There are many ways of defining the integrated particle current in the
system: one can consider the current through a given bond, take a
space-averaged current or measure the distance travelled by a tagged
particle (in periodic systems). All these definitions give a different
current for finite times but in the $t\to\infty$ limit they are essentially
the same, since the difference between them is bounded while they
are proportional to $t$.

However, for systems with infinite configuration space this argument
does not hold and the boundary terms can be relevant even for large
times, leading to several independent relations for the fluctuations. 

The fact that the difference between two currents can be written as
a boundary term suggests that the same holds for the difference between
two initial states. Suppose that the transient fluctuation theorem
holds for $J^{(1)},E^{(1)}$ with the initial measure $p_{\sigma}^{\mathrm{eq(1)}}=\rme^{-V_{\sigma}^{(1)}}$.
Since for any two currents $E^{(1)}J^{(1)}=E^{(2)}J^{(2)}+V_{\sigma_{\mathrm{ini}}}^{(2)}-V_{\sigma_{\mathrm{fin}}}^{(2)}-V_{\sigma_{\mathrm{ini}}}^{(1)}+V_{\sigma_{\mathrm{fin}}}^{(1)}$,
the transient fluctuation theorem holds for any current $J$ with
a boundary term as 
\begin{equation} 
J_{\mathrm{corr}}=EJ+V_{\sigma_{\mathrm{ini}}}-V_{\sigma_{\mathrm{fin}}}-V_{\sigma_{\mathrm{ini}}}^{\mathrm{ini}}+V_{\sigma_{\mathrm{fin}}}^{\mathrm{ini}},\label{eq:Jcorrected}
\end{equation}
 where $V$ is the equilibrium potential corresponding to the current
$J$ and $V_{\sigma}^{\mathrm{ini}}=\ln p_{\sigma}^{\mathrm{ini}}$
with $p_{\sigma}^{\mathrm{ini}}$ being the initial measure ($J_{\mathrm{corr}}$
satisfies the theorem with $E=1$).

\subsection{Breakdown of the fluctuation theorem in infinite systems}

For infinite systems the above argument breaks down at (\ref{eq:as}),
where one identifies $e(\lambda)$ defined in (\ref{eq:elambda})
with the lowest eigenvalue of $\tilde{H}(\lambda)$. In cases where
$\tilde{H}$ is infinite-dimensional, there is no guarantee that the
scalar products appearing in (\ref{eq:as}) are finite. An explicit
example of this breakdown is given in \cite{Harris2006} and will be
discussed in detail below.  Note however, that the derivation of the transient fluctuation theorem
still holds even for infinite systems. Consequently, a violation of
the asymptotic fluctuation theorem requires the unbounded growth of
the boundary term in (\ref{eq:Jcorrected}). 

To avoid possible confusion, we remark here that in the
original dynamical systems formulation \cite{Gallavotti95,Gallavotti95b} of the GC
symmetry, a critical value $r^*$ already appears above which the relation
(\ref{e:GCFT}) does not hold.  This is a direct consequence of the fact that the
large deviation function (equivalent to $\hat e$ in our notation) diverges outside a
finite interval.  In contrast, in the stochastic framework of this
paper, we consider situations where the large deviation function is always
finite (i.e., formally $r^*=\infty$) but the symmetry may still have a
restricted range of validity due to the relevance of boundary terms.
Indeed, it has also recently been understood in the dynamical
systems context, that adding singular boundary terms to a
system with finite $r^*$ leads to a further reduction in the
symmetry regime, see~\cite{Bonetto06b}.

\subsection{Extended fluctuation theorem}
\label{ss:EFT}

The argument we present in this section is largely based on that of van Zon and
Cohen \cite{VanZon03,VanZon04}, where they describe a generic scenario
for the breakdown of the fluctuation theorem in the context of
Langevin dynamics.   This leads to an ``extended fluctuation
theorem'', where the quantity $\hat e(j)-\hat e(-j)$ is linear for
small $j$ (as suggested by the fluctuation theorem) but, after an
intermediate crossover regime, saturates to a constant value for large
$j$.   Similar arguments are also given in~\cite{Baiesi06}.

Our starting point is equation (\ref{eq:Jcorrected}) which we now write for the time-intensive quantities:
\begin{equation}
 j_\mathrm{corr} = j + t^{-1}(B_\mathrm{ini} - B_\mathrm{fin}).
\end{equation} 
Here $j_\mathrm{corr}$ is the corrected current for which the
fluctuation theorem holds, and $B$ is a boundary term, which depends only on the initial/final configuration of the history.

First we assume that the probability distribution $\mathcal{P}_{(B)}$ of $B$ has an exponential tail for very large/small values of $B$:
\begin{equation}
 \mathcal{P}_{(B)}(B,t)\sim \left\{ 
\begin{array}{ll} 
\rme^{-{A^+ B}} & \quad B\to\infty \\ 
\rme^{{A^- B}} & \quad B\to-\infty
\end{array}
\right..
\end{equation} 
Here we set $A^{+/-}$ to infinity if $B$ is bounded from above/below. 
Focusing on the case where the initial state is the steady state, one can assume that the probability distribution of $B$ is identical for the initial and final state, moreover, for large measurement times they become independent. This implies that the probability distribution of 
\begin{equation}
 b=t^{-1}(B_\mathrm{ini} - B_\mathrm{fin})
\end{equation}  
is of the form
\begin{equation}
 \mathcal{P}_{(b)}(b,t) \sim \rme^{-tA|b|},
\end{equation} 
where $A=\min(A^+,A^-)$.

\begin{figure}
\begin{center}
 \include{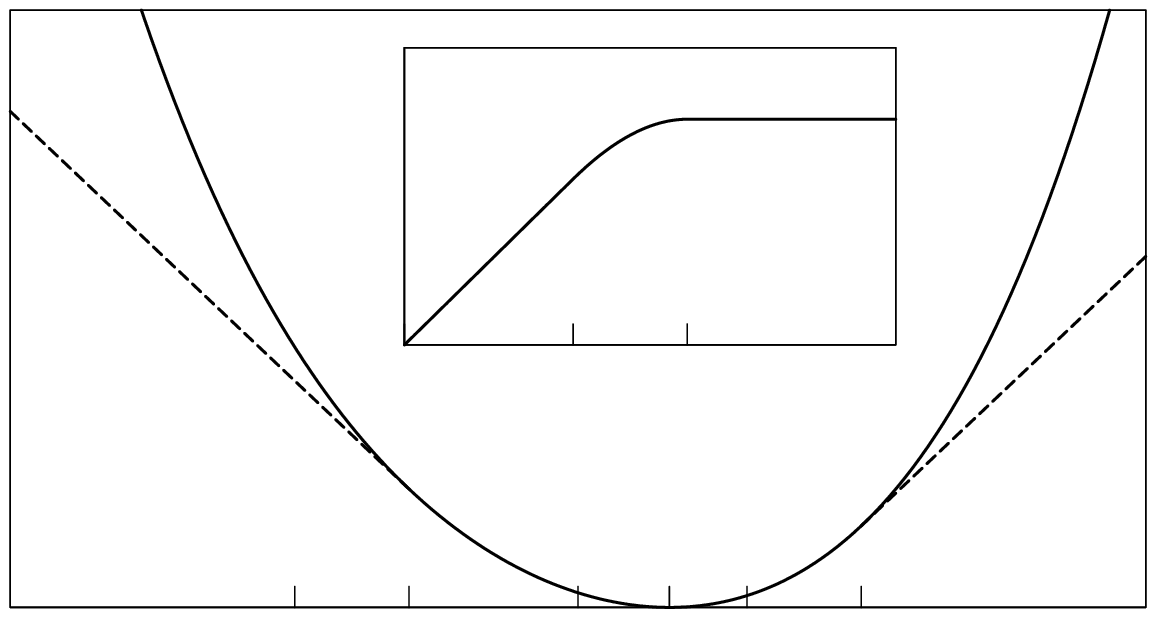}
\end{center}
\caption{A schematic plot showing the large deviation functions $e(j)$ and $\hat{e}_\mathrm{c}(j)$. While the fluctuation theorem holds for $e_\mathrm{c}(j)$, a modified form is found for $\hat{e}(j)$, as plotted in the inset}
\label{fig:extended}
\end{figure}

As a next step we assume that $b$ and $j_\mathrm{corr}$ as random variables are independent. In this case the large deviation function of the fluctuations of $j=j_\mathrm{corr}+b$ are determined by
\begin{equation} \label{eq:ej}
 \hat e(j) = \min_{j_\mathrm{corr}} (\hat e_\mathrm{c}(j_\mathrm{corr})+ A|j_\mathrm{corr}-j|).
\end{equation} 
Here $\hat e_\mathrm{c}$, which describes the large deviations of $j_\mathrm{corr}$, satisfies the fluctuation theorem, i.e., $\hat e_\mathrm{c}(-j)-\hat e_\mathrm{c}(j)=j$. In figure \ref{fig:extended}, one can see how  $\hat e(j)$ can be obtained from $\hat e_\mathrm{c}(j)$ graphically by using (\ref{eq:ej}). The points $-j_1$ and $j_2$ are defined as those values of $j$ where the derivative of the convex function $\hat e_\mathrm{c}(j)$ becomes $-A$ and $A$ respectively. Outside the interval $(-j_1,j_2)$ the function $\hat e(j)$ is linear with slope $\pm A$. Consequently, the quantity $\hat e(-j)-\hat e(j)$ is linear from zero to $j=\min(j_1,j_2)$, and after a crossover regime it saturates at $j=\max(j_1,j_2)$ to a constant value $C$. We remark that in the case where $e_\mathrm{c}(j)$ is symmetric with respect to $\langle j\rangle$ (e.g. in the case of Gaussian fluctuations) $C=2A\langle j\rangle$.

In a more general situation, where the initial state is not stationary, the fluctuations of $B$ are different in the initial and final state (we still require that the tails are exponential). 
\begin{equation}
  \mathcal{P} _{(B_\mathrm{ini})}(B,t)\sim \left\{ 
\begin{array}{ll} 
\rme^{-{A_\mathrm{ini}^+ B}} & \quad B\to\infty \\ 
\rme^{{A_\mathrm{ini}^- B}} & \quad B\to-\infty
\end{array}
\right. .
\end{equation}  
In this case the large deviations of $b$ take the following form:
\begin{equation}
  \mathcal{P}_{(b)}(b,t) \sim \left\{\begin{array}{ll}
                   \rme^{-tA_1b} \quad & b>0 \\
                   \rme^{tA_2b} & b<0 
                  \end{array}\right.,
\end{equation} 
where $A_1=\min(A^+_\mathrm{ini}, A^-)$ and $A_2=\min(A^-_\mathrm{ini}, A^+)$. This leads to a variant of the extended fluctuation theorem, where the the quantity $\hat{e}(-j)-\hat{e}(j)$ does not saturate but becomes linear with a non-zero slope for large $j$. 

We stress that the above argument relies strongly on the assumption
that the boundary terms are independent of the bulk
contribution. There are various examples in the literature where this
holds, and the above extended form of the fluctuation theorem was
found, e.g., \cite{VanZon04,VanZon03,Puglisi06}. In contrast, in
the remainder of this paper, we discuss a simple stochastic Markovian
system where there is a strong correlation between the boundary and
bulk terms and hence this extended fluctuation theorem is not expected to hold.

\section{Zero-range process}

\label{s:ZRP}

Here we demonstrate how the general formalism of the preceding section
is applied to a specific model---the zero-range process (ZRP). First
introduced by Spitzer in 1970~\cite{Spitzer70}, the ZRP now plays
a paradigmatic role in non-equilibrium statistical mechanics, see~\cite{Evans05}
for a recent review. In particular, for certain choices of parameters
the model exhibits a condensation transition~\cite{Evans00,Jeon00b}
in which a macroscopic proportion of particles pile up on a single
site. Condensation phenomena are well-known in colloidal and granular
systems~\cite{Shim04} and also occur in a variety of other physical
and non-physical contexts~\cite{Evans05}.

\subsection{Model}
\begin{figure}
\psfrag{1}[][]{1}
\psfrag{2}[][]{2}
\psfrag{3}[][]{3}
\psfrag{4}[][]{4}
\psfrag{5}[][]{5}
\psfrag{L-2}[][]{$L\!-\!2$}
\psfrag{L-1}[][]{$L\!-\!1$}
\psfrag{L}[][]{$L$}
\psfrag{a}[][]{$\alpha$}
\psfrag{d}[][]{$\delta$}
\psfrag{bw}[][]{$\beta w_n$}
\psfrag{cw}[][]{$\gamma w_n$}
\psfrag{I}[][]{``in''}
\psfrag{O}[][]{``out''}
\psfrag{pw}[][]{$p w_n$}
\psfrag{qw}[][]{$q w_n$}
\begin{centering}
\includegraphics[width=10cm]{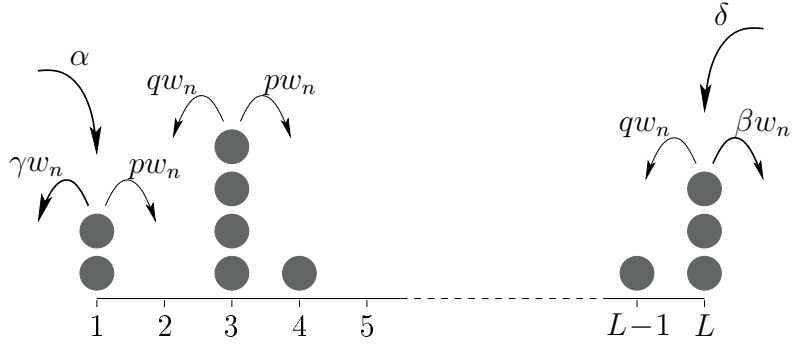} \par
\end{centering}
\caption{Schematic representation of the ZRP on an open $L$-site lattice}
\label{f:ZRP} 
\end{figure}

We study the one-dimensional partially asymmetric
zero-range process with open boundaries~\cite{Levine04c}---see
figure~\ref{f:ZRP}. %
Each lattice site can be occupied by any integer number of particles,
the uppermost of which hops randomly to a nearest neighbour site after
an exponentially distributed waiting time. In the bulk particles move
to the right (left) with rate $pw_{n}$ ($qw_{n}$) where $w_{n}$
is a function of the number of particles $n$ on the departure site
($w_{0}=0$ by definition). Note that $w_{n}=n$ would correspond
to free particles whereas all other forms represent an attractive
or repulsive inter-particle interaction. The top particle on the leftmost
lattice site (site 1) leaves the system with rate $\gamma w_{n}$
whereas new particles are injected with rate $\alpha$. Correspondingly,
on the rightmost site (site $L$) particles are removed (injected)
with rates $\beta w_{n}$ ($\delta$). For later convenience we label
each bond by the site at its left-hand end, i.e., the $l$th bond
is between sites $l$ and $l+1$.

In the quantum Hamiltonian formalism this dynamics
is encoded in the Hamiltonian  
\begin{eqnarray}
\fl H=-\biggl\{\sum_{l=1}^{L-1}\left[p(a_{l}^{-}a_{l+1}^{+}-d_{l})+q(a_{l}^{+}a_{l+1}^{-}-d_{l+1})\right]\cr
+\alpha(a_{1}^{+}-1)+\gamma(a_{1}^{-}-d_{1})+\delta(a_{L}^{+}-1)+\beta(a_{L}^{-}-d_{L})\biggr\}\label{e:H}
\end{eqnarray}
 where $a^{+}$ and $a^{-}$ are infinite-dimensional particle creation
and annihilation matrices 
\begin{equation} \fl
a^{+}=\left(\begin{array}{ccccc}
0 & 0 & 0 & 0 & \cdots\\
1 & 0 & 0 & 0 & \cdots\\
0 & 1 & 0 & 0 & \cdots\\
0 & 0 & 1 & 0 & \cdots\\
\vdots & \vdots & \vdots & \vdots & \ddots\end{array}\right),\qquad a^{-}=\left(\begin{array}{ccccc}
0 & w_{1} & 0 & 0 & \cdots\\
0 & 0 & w_{2} & 0 & \cdots\\
0 & 0 & 0 & w_{3} & \cdots\\
0 & 0 & 0 & 0 & \cdots\\
\vdots & \vdots & \vdots & \vdots & \ddots\end{array}\right)\label{eq:aa+}
\end{equation}
 and $d$ is a diagonal matrix with the $(i,j)$th element given by
$w_{i}\delta_{i,j}$.

\subsection{Stationary state, current fluctuations}
\label{ss:stationary_state}

The steady state of the ZRP is given by a product measure (in other
words, the stationary distributions for each site are uncorrelated):
\begin{equation}
|P^{*}\rangle=|P_{1}^{*})\otimes|P_{2}^{*})\otimes\ldots\otimes|P_{L}^{*})\label{e:prod}
\end{equation}
 where $|P_{l}^{*})$ is the probability vector with components 
\begin{equation}
P^{*}(n_{l}=n)=\frac{z_{l}^{n}}{Z_{l}}\prod_{i=1}^{n}w_{i}^{-1}\label{e:prob}.
\end{equation}
Here the empty product $n=0$ is defined equal to 1 and $Z_{l}$ is
the local analogue of the grand-canonical partition function 
\begin{equation}
Z_{l}\equiv Z(z_{l})=\sum_{n=0}^{\infty}z_{l}^{n}\prod_{i=1}^{n}w_{i}^{-1}. \label{e:gc}
\end{equation}
 The fugacities $z_{l}$ are uniquely determined by the hopping rates
$\alpha$, $\beta$, $\gamma$, $\delta$, $p$ and $q$. However,
for $w_{n}$ bounded (i.e., $\lim_{n\to\infty}w_{n}=a$ with $a<\infty$),
$Z_{l}$ has a finite radius of convergence. For parameters
leading to fugacities outside this radius of convergence, a growing
boundary condensate occurs~\cite{Levine04c}.

We are interested in the application of the fluctuation theorem to
this model, for the case where the boundary parameters are chosen
to give a well-defined steady state, i.e., without boundary condensation. As discussed in the preceding
section, one can consider a variety of different {}``currents''
through the system. However, since the state space is unbounded (each
lattice site can contain an arbitrarily large number of particles)
we anticipate the possibility of relevant boundary terms and different
fluctuation relationships.

A natural choice is to look at the physical current of particles
across a particular bond.  For example, suppose we choose to focus on the particle current into
the system (i.e., across the $0$th bond), then the modified Hamiltonian is given by \begin{eqnarray} \fl
\tilde{H}(\lambda)=-\biggl\{\sum_{l=1}^{L-1}\left[p(a_{l}^{-}a_{l+1}^{+}-d_{l})+q(a_{l}^{+}a_{l+1}^{-}-d_{l+1})\right]\nonumber \\
+\alpha(a_{1}^{+}\rme^{-\lambda}-1)+\gamma(a_{1}^{-}\rme^{\lambda}-d_{1})+\delta(a_{L}^{+}-1)+\beta(a_{L}^{-}-d_{L})\biggr\}.\label{e:H0}
\end{eqnarray}
In the next section we will examine closely the spectrum of this
Hamiltonian for the $L=1$ single site case. Here we recap some known
results for the general $L$-site case as obtained in~\cite{Me05}.

\emph{If $w_{n}$ is unbounded}, (i.e., $\lim_{n\to\infty}w_{n}=\infty$)
then $\tilde{H}(\lambda)$ has a gapped spectrum for all $\lambda$ with lowest
eigenvalue given by\footnote{The leading order term in an $L \to \infty$ expansion for the bulk-symmetric case,
  corresponding to the limit $p=q=1$ in~(\ref{e:elamb}), has also been
  obtained by an additivity principle~\cite{Bodineau04} and by field-theoretic methods~\cite{Wijland04}.}
\begin{equation}
\epsilon_{0}(\lambda)=\frac{(p-q)(\rme^{\lambda}-1)\left[\alpha\beta\left(\frac{p}{q}\right)^{L-1}\rme^{-\lambda}-\gamma\delta\right]}{\gamma(p-q-\beta)+\beta(p-q+\gamma)\left(\frac{p}{q}\right)^{L-1}}.\label{e:elamb}
\end{equation}
The right eigenvector $|\psi_0\rangle$ corresponding to (\ref{e:elamb})
has the same form as the stationary product measure (\ref{e:prod})
with fugacities
\begin{equation} \fl
z_{l}=\frac{[(\alpha \rme^{-\lambda}+\delta)(p-q)-\alpha\beta \rme^{-\lambda}+\gamma\delta]\left(\frac{p}{q}\right)^{l-1}-\gamma\delta+\alpha\beta \rme^{-\lambda}\left(\frac{p}{q}\right)^{L-1}}{\gamma(p-q-\beta)+\beta(p-q+\gamma)\left(\frac{p}{q}\right)^{L-1}}.\label{e:fug}
\end{equation}
 and the left-hand eigenvector is also a product state with one-site marginal ``fugacities''
\begin{equation} \fl
\tilde{z}_{l}=\frac{\beta\gamma(\rme^{\lambda}-1)\left(\frac{p}{q}\right)^{L-l}+\gamma \rme^{\lambda}(p-q-\beta)+\beta(p-q+\gamma)\left(\frac{p}{q}\right)^{L-1}}{\gamma(p-q-\beta)+\beta(p-q+\gamma)\left(\frac{p}{q}\right)^{L-1}}.\label{e:lfug}
\end{equation}
 Note that this result is independent of the details of $w_{n}$ and
one can also show that it is the same for currents across all bonds
meaning that the current fluctuations are spatially homogeneous in
the long-time limit. Furthermore the scalar products appearing in
equation (\ref{eq:as}) are finite and the standard Gallavotti-Cohen
fluctuation theorem is recovered. Physically, unbounded $w_{n}$ means
that there is no chance for the temporary accumulation of a large
numbers of particles on any site.

However, \emph{if $w_{n}$ is bounded}, then there is a crossover to a
  gapless spectrum at some value of $\lambda$ and also the scalar products
in equation (\ref{eq:as}) can be infinite.  In fact,
  $\langle\psi_0|P_{0}\rangle$ and $\langle s|\psi_0 \rangle$ are related to
  the distribution of initial and final boundary terms and the condition
for $\langle\psi_0|P_{0}\rangle$ to diverge obviously depends on the initial state.
Thus, although the symmetry of the eigenvalues remains, we expect
the breakdown of the usual Gallavotti-Cohen fluctuation theorem in a spatially
homogeneous and initial-condition-dependent way. To explore this issue
in more detail we present, in the next section, explicit calculations
for the single-site case. 

\section{Analytical results for single-site model}

\label{s:calc}

\begin{figure}
\begin{center}
\psfrag{l}[][]{l}
\psfrag{a}[][]{$\alpha$}
\psfrag{d}[][]{$\delta$}
\psfrag{bw}[][]{$\beta$}
\psfrag{cw}[][]{$\gamma$}
\psfrag{I}[][]{``in''}
\psfrag{O}[][]{``out''}
\includegraphics*[width=0.165\textwidth]{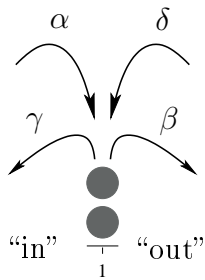}
\caption{Schematic representation of the single-site ZRP with $w_n=1$.\label{f:singsite}}
\end{center}
\end{figure}

As a simple example of the ZRP with bounded $w_{n}$, we now focus on 
the single-site model with $w_{n}=1$---see figure~\ref{f:singsite}.
Even this apparently simple model exhibits a rich phase behaviour,
as shown in~\cite{Harris2006}.
One must now distinguish between the particle currents across only two bonds
viz. the 0th ({}``input'') and 1st ({}``output'').  Since fluctuations
across the two bonds are
simply related by left-right reflection, we will consider only the
former except where explicitly stated otherwise. Note that the occupation number $n$
of the site performs a random walk on a semi-infinite lattice
but with two independent processes for movement in each direction.
For a well-defined steady state this random walk should be biased
towards the reflecting boundary at $n=0$, i.e., we require $\alpha+\delta<\beta+\gamma$.
If this condition is not met then there is a growing condensate on
the site.

Let us first consider an initial Boltzmann distribution 
given by 
\begin{equation}
|P_{0}\rangle=(1-x)\sum_{n=0}^{\infty}x^{n}|n\rangle\label{e:boltz}
\end{equation}
 where $|n\rangle$ denotes the configuration with site occupied by $n$ particles
(i.e., a column vector with a `1' in the $n$th position and `0's
elsewhere) and the {}``fugacity'' $x$ is less than 1 for normalization.
For example, the choice $x=(\alpha+\delta)/(\beta+\gamma)$ is the
steady-state initial condition. Note also that the limit
$x\to0$ corresponds to the empty-site case and that by ergodicity
this gives the same result as any fixed particle configuration.

To obtain the large deviations of input current, we need to calculate
the matrix element $\langle s|\rme^{-\tilde{H}t}|P_{0}\rangle$ where
the modified Hamiltonian $\tilde{H}(\lambda)$ is given by 
\begin{equation} \fl
\left( \begin{array}{ccccc}\alpha+\delta & -\gamma \rme^{\lambda}-\beta & 0 & 0 & \ldots\\
-\alpha \rme^{-\lambda}-\delta & \alpha+\beta+\gamma+\delta & -\gamma \rme^{\lambda}-\beta & 0 & \ldots\\
0 & -\alpha \rme^{-\lambda}-\delta & \alpha+\beta+\gamma+\delta & -\gamma \rme^{\lambda}-\beta & \ldots\\
0 & 0 & -\alpha \rme^{-\lambda}-\delta & \alpha+\beta+\gamma+\delta & \ldots\\
\vdots & \vdots & \vdots & \vdots & \ddots \end{array}\right) \label{eq:tildeH}
\end{equation}
In appendix~\ref{ss:spectrum}, we
first present an explicit calculation of the spectrum of $\tilde{H}(\lambda)$ and thereby illustrate
the crossover to a gapless regime. Then,
in appendix~\ref{ss:ld}, we show how to obtain an integral representation
of the matrix element $\langle s|\rme^{\tilde{H}(\lambda)}|P_{0}\rangle$
and thus extract the large deviation behaviour. This enables us to
construct and explain the ``phase diagram'' for current fluctuations
as shown in section~\ref{ss:phase}. Finally, in~\ref{ss:gen} we discuss
some straightforward generalizations of these results.

\subsection{Phase diagrams}
\label{s:phase_diagrams}

In order to extract the large-time behaviour from the integral representation
(\ref{eq:final}) we can use the method of steepest descents with
saddle-point at $z=1$. However, care must be taken due to the poles
in the integrands (at $z=\phi^{-1}$, $(\phi x)^{-1}$,$\phi$ and
$y$). If the saddle-point contour has to be deformed through one
of these poles then we must take into account the contribution of
the residue at that pole. For a given $\lambda$, $e(\lambda)$ is
then determined by the term which decays most slowly with $t$. This
yields changes in behaviour at the values of $\lambda$ given in table~\ref{t:lam},
where we have defined for convenience the parameter combination 
\begin{equation}
\eta=\sqrt{[(\beta+\gamma)^{2}-\beta\delta-\alpha\gamma]^{2}-4\alpha\beta\gamma\delta}.
\end{equation}
 
\begin{table}
\begin{center}
\begin{tabular}{ll}
\small{Values of $\lambda$} &
\small{Corresponding values of $j$} \tabularnewline 
$\rme^{\lambda_{1}}\equiv\frac{\alpha}{\beta+\gamma-\delta}$ &
$j_{a}\equiv\frac{(\beta+\gamma-\delta)^{2}-\alpha\gamma}{\beta+\gamma-\delta}$,
$j_{b}\equiv\frac{\beta(\beta+\gamma-\delta)^{2}-\alpha\gamma\delta}{(\beta+\gamma)(\beta+\gamma-\delta)}$ \tabularnewline 
$\rme^{\lambda_{2}}\equiv\frac{(\beta+\gamma)^{2}-\alpha\gamma-\beta\delta+\eta}{2\gamma\delta}$ &
$j_{c}\equiv-\frac{\eta}{\beta+\gamma}$ \tabularnewline
$\rme^{\lambda_{3}}\equiv\frac{\delta-\beta x^{2}+\sqrt{(\delta-\beta x^{2})^{2}+4\alpha\gamma x^{2}}}{2\gamma x^{2}}$ &
$j_{d}\equiv\frac{-(\delta-\beta x^{2})}{x}$ \tabularnewline 
$\rme^{\lambda_{4}}\equiv\frac{\beta(1-x)+\gamma}{\gamma x}$ &
$j_{e}\equiv\frac{\alpha\beta\gamma x^{2}-\delta\left[\beta(1-x)+\gamma\right]^{2}}{x(\beta+\gamma)\left[\beta(1-x)+\gamma\right]}$,
$j_{f}\equiv\frac{\alpha\gamma-\left[\beta(1-x)+\gamma\right]^{2}}{\beta(1-x)+\gamma}$
\end{tabular}
\end{center}
\caption{\label{t:lam} Transition values for input current fluctuations in
single-site ZRP.}
\end{table}

\label{ss:phase}

\begin{figure}
\psfrag{x}[][]{$x$}
\psfrag{A}[][]{A}
\psfrag{B}[][]{B}
\psfrag{C}[][]{C}
\psfrag{D}[][]{D}
\psfrag{l}[][]{$\lambda$}
\psfrag{-2}[][]{-2}
\psfrag{-1}[][]{-1}
\psfrag{0}[][]{0}
\psfrag{1}[][]{1}
\psfrag{2}[][]{2}
\psfrag{3}[][]{3}
\psfrag{4}[][]{4}
\psfrag{5}[][]{5}
\psfrag{0.25}[][]{0.25}
\psfrag{0.5}[][]{0.5}
\psfrag{0.75}[][]{0.75}
\psfrag{xc}[Tc][Bc]{$x_c$}
\begin{centering}
\includegraphics[width=7cm]{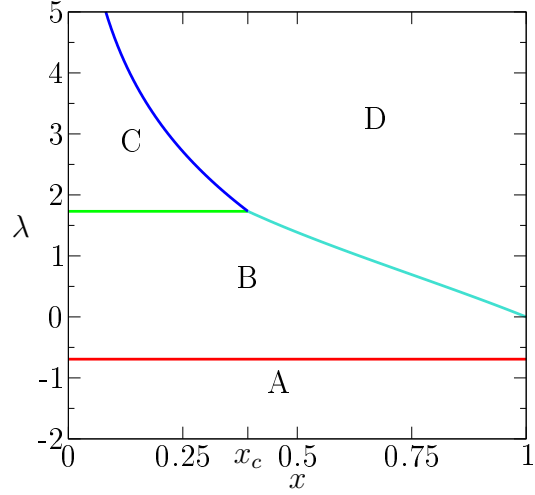}\par
\end{centering}
\caption{The $x$--$\lambda$ phase diagram for $\alpha=\gamma=\delta=0.1$,
$\beta=0.2$. Red line: $\lambda=\lambda_{1}$, green line: $\lambda=\lambda_{2}$,
blue line: $\lambda=\lambda_{3}(x)$, cyan line:
$\lambda=\lambda_{4}(x)$.
\label{fig:mu-lambda}}
\end{figure}

\begin{lyxlist}{00.00.0000}
\item [{Phase~A:}] $\lambda<\lambda_{1}$ (see figure \ref{fig:mu-lambda}).
Here $e(\lambda)$ takes the form 
\begin{equation}
e(\lambda)=\alpha\left(1-\rme^{-\lambda}\right)+\gamma\left(1-\rme^{\lambda}\right),
\end{equation}
which does not coincide with the lowest eigenvalue of $\tilde{H}$.
The reason is that here the quantity $\left\langle s|\psi_{0}\right\rangle $
in (\ref{eq:as}) diverges. This phase corresponds to large forward
currents.
\item [{Phase~B:}] $\left[\left(x<x_{c}\right)\land\left(\lambda_{1}<\lambda<\lambda_{2}\right)\right]\lor\left[\left(x> x_{c}\right)\land\left(\lambda_{1}<\lambda<\lambda_{4}\right)\right]$,
where we defined $x_{c}$ as 
\begin{equation}
x_{c}=\frac{-\eta+(\beta+\gamma)^{2}-\alpha\gamma+\beta\delta}{2\beta(\beta+\gamma)}
\end{equation}
 (see figure \ref{fig:mu-lambda}). In this region the spectrum of
$\tilde{H}$ is gapped and $e(\lambda)$ coincides with the lowest
eigenvalue:
\begin{equation}
e(\lambda)
=\frac{\alpha\beta}{\beta+\gamma}(1-\rme^{-\lambda})+\frac{\gamma\delta}{\beta+\gamma}(1-\rme^\lambda).
\end{equation}
The quantities $\left\langle s|\psi_{0}\right\rangle $ and $\left\langle \psi_{0}|P_{0}\right\rangle $
are finite. 
\item [{Phase~C:}] $\left(x<x_{c}\right)\land\left(\lambda_{2}<\lambda<\lambda_{3}\right)$
(see figure \ref{fig:mu-lambda}). In this phase the spectrum of $\tilde{H}$ is gapless. The
scalar products $\langle s|\psi_{0}\rangle$ and $\langle\psi_{0}|P_{0}\rangle$
are finite (here $|\psi_{0}\rangle$, which is the ground-state of $\tilde{H}$, has to be understood as $|\psi_{k\to 0}\rangle$), consequently $e(\lambda)$ is given by the lowest eigenvalue:
\begin{equation}
e(\lambda)=\alpha+\beta+\gamma+\delta-2\sqrt{\left(\alpha \rme^{-\lambda}+\delta\right)\left(\beta+\gamma \rme^{\lambda}\right)}
\end{equation}
\item [{Phase~D:}] $\left[\left(x<x_{c}\right)\land\left(\lambda>\lambda_{3}\right)\right]\lor\left[\left(x>x_{c}\right)\land\left(\lambda>\lambda_{4}\right)\right]$
(see figure \ref{fig:mu-lambda}). In this phase the quantity $\left\langle \psi_{0}|P_{0}\right\rangle $
diverges and $e(\lambda)$ differs from the lowest eigenvalue of $\tilde{H}(\lambda)$:
\begin{equation}
e(\lambda)
=\alpha+\beta+\gamma+\delta-\left(\beta+\gamma \rme^{\lambda}\right)x-\frac{\alpha \rme^{-\lambda}+\delta}{x}.
\end{equation}
 It is interesting that here the large deviation of current fluctuations
retains a dependence on the initial state of the system.
\end{lyxlist}

\begin{figure}
\psfrag{I}[][]{A} 
\psfrag{II}[][]{B} 
\psfrag{III}[][]{C} 
\psfrag{IV}[][]{D} 
\psfrag{m}[][]{$x$} 
\psfrag{j }[][]{$j$} 
\psfrag{0}[][]{0} 
\psfrag{0.1}[][]{0.1} 
\psfrag{0.2}[][]{0.2} 
\psfrag{0.4}[][]{0.4} 
\psfrag{0.6}[][]{0.6} 
\psfrag{0.8}[][]{0.8} 
\psfrag{1}[][]{1} 
\psfrag{-0.1}[][]{-0.1} 
\psfrag{-0.2}[][]{-0.2} 
\psfrag{-0.3}[][]{-0.3} 
\psfrag{-0.4}[][]{-0.4} 
\psfrag{-0.5}[][]{-0.5} 
\begin{center}
 \includegraphics[width=9cm,bb=   5   200   549   583]{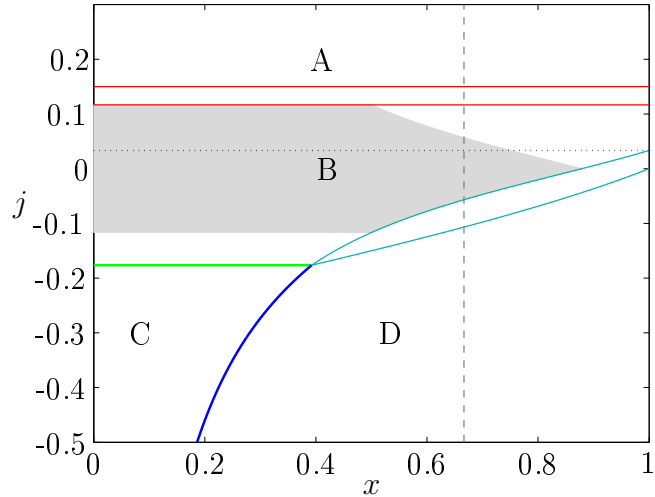}
\end{center}
\caption{The $x$--$j$ phase diagram for $\alpha=\gamma=\delta=0.1$,
$\beta=0.2$. At first order transition lines (A--B and B--D) mixed phases appear. Red line: $j=j_a$ or $j=j_b$, green line: $j=j_c$, blue line: $j=j_d(x)$, cyan line: $j=j_e(x)$ or $j=j_f(x)$. The horizontal dotted line shows the mean current, the vertical dashed line indicates the the specific value of $x$ which corresponds to the steady state initial condition.\label{fig:mu-j}}
\end{figure}

For the physical interpretation of these phases it is better to consider
$\hat{e}(j)$, which can be obtained from $e(\lambda)$ by a Legendre
transformation. $\hat{e}(j)$ has the following forms in the
different regions of figure~\ref{fig:mu-j}: 
\begin{equation}
\hat{e}(j)=\left\lbrace \begin{array}{ll}
f_{j}(\alpha,\gamma) & \mathrm{A}\\
f_{j}\left(\frac{\alpha\beta}{\beta+\gamma},\frac{\gamma\delta}{\beta+\gamma}\right) & \mathrm{B}\\
f_{j}(\alpha,\gamma)+f_{j}(\beta,\delta) & \mathrm{C}\\
f_{j}(\alpha,\gamma)+\beta(1-x)+\delta\left(1-x^{-1}\right)+j\ln x & \mathrm{D}\end{array}\right. \label{e:ejres}
\end{equation}
 with 
\begin{equation}
f_{j}(a,b) =a+b-\sqrt{j^{2}+4ab}+j\ln\frac{j+\sqrt{j^{2}+4ab}}{2a}.\label{e:fj}
\end{equation}
We remark that $f_{j}(a,b)$ has a simple physical meaning. Consider
a single particle performing a simple random walk on the infinite
one-dimensional lattice with right (left) hopping rate $a$ ($b$).
Then the large deviation function of the distance travelled by this
particle is given by $f_{j}(a,b)$. 

In phase A only the rates for the first bond determine the large deviation
function. For such large currents particles typically pile up on the
site, which then acts as an infinite reservoir. In this case the behaviour
of the two bonds decouples, so the current distribution across the
input bond is entirely controlled by the two Poisson processes at
rate $\alpha$ and $\gamma$. This explains the first line in (\ref{e:ejres}).
The fact that in these (very unlikely) realizations the occupation
number increases with time, is consistent with the divergence of $\langle s|\psi_0\rangle$.
In stochastic systems such divergence is usually associated with the
lack of a steady state. 
Here, the model does have a steady state so that realizations
involving piling-up of particles (condensation) are never observed in
the infinite time limit.  However, as indicated, they characterize the
fluctuations that can be observed over a large but finite measuring
period $t$.  We therefore use the term ``instantanous condensate''.

In phase B, which always contains the mean steady-state current ($\lambda=0$),
the four rates enter symmetrically in the large deviation function.
Note that the combinations $\alpha\beta/(\beta+\gamma)$ and $\gamma\delta/(\beta+\gamma)$
are the effective renormalized hopping rates of two exclusion particles
with left (right) hopping rates $\alpha$ ($\gamma$) and $\beta$
($\delta$), in the case where they form a bound state \cite{Juhasz2005}.
In this regime the occupation of the site, which maps to the distance
between the two exclusion particles, remains finite. The existence
of this {}``two-particle bound state'' is also manifested in the
gapped spectrum of $\tilde{H}$. 

Although the naive approach (i.e. identifying $e(\lambda)$ with the
lowest eigenvalue of $\tilde{H}$) still works in phase C, the large
deviation function takes a different form here. For these large negative
currents (and small initial occupation) it is needed that the current
across the other bond also takes a large negative value (unlike in
phase A). It can be seen in (\ref{e:ejres}) that the contribution
of the two bonds factorize suggesting that they act independently.
This can be traced back to the change in the spectrum which becomes
gapless in this regime, corresponding to a two-particle unbound state. 
The distance between the two particles (or equivalently the occupation
of the site) grows as the square-root of time in such
realizations---again we refer to this temporary piling-up as
instantaneous condensation.

Phase D is in some sense the counterpart of phase A. Typical realizations
contributing to these exponentially small probabilities start at a
very high occupation (initial condensate) which then decreases during
the observation time. For fixed $j$, this is possible only for sufficiently large
values of $x$. This initial singularity is indicated by the divergence
of $\langle\psi_0|P_{0}\rangle$.

One can see in the $x$--$j$ phase diagram (see figure \ref{fig:mu-j}) 
that in between phases
A--B and B--D transition regions appear. These regimes correspond
to a single transition line in the $x$--$\lambda$ diagram along
which the derivative of $e(\lambda)$ is discontinuous. This is entirely
analogous to ordinary equilibrium phase transitions where different thermodynamical
potentials are Legendre transforms of each other and in certain phase
diagrams mixed regions (e.g. liquid--gas) appear at first order phase
transitions lines. For the full analogy one can consider the following
identifications: $j\to$(specific) volume; $\lambda\to$pressure;
$\hat{e}(j)\to$ Helmholtz free energy (density); $e(\lambda)\to$
Gibbs free energy (density). One can immediately see that the analogue
quantity of the system size $N$ is the measurement time $t$ in our
model which must diverge if a true phase transition is to exist.
In the mixed regions of the phase diagram the system segregates in
time, i.e., for some finite fraction of the whole measurement time
the system behaves as being on one boundary of the mixed phase while
for the rest of the time it switches to the other
boundary.\footnote{Mathematically speaking, knowledge of $e(\lambda)$
  is only sufficient to determine $\hat{e}(j)$ where $e(\lambda)$ is
  differentiable.  If $e(\lambda)$ is non-differentiable then
  $\hat{e}(j)$ is, in general, non-convex; Legendre transform of the
  non-differentiable points in $e(\lambda)$ then yields straight-line
  sections of the convex envelope of $\hat{e}(j)$.   However in our
  case, the physical argument based on phase separation in time,
  indicates that the linear sections obtained via Legendre transform
  \emph{do} yield the correct form for $\hat e(j)$ in the transition
  regimes (i.e., the large deviation function \emph{is} still
  convex). For a mathematical account of the subtleties of large
  deviation theory the reader is referred to~\cite{Dembo98};
  the application to statistical mechanics is discussed in~\cite{Oono89,PreTouchette2008}.}

One can formally consider $t$ as a space dimension. Then a path in
the configuration space of the original model becomes a configuration
of the new model, see e.g.,~\cite{Lecomte07c}.   If the original model has finite number of configurations then this leads to only one (diverging) dimension ($t$) in the new model hence no phase transition is possible in this case. Therefore for a real phase transition to take place one needs infinitely many possible configurations in the original model. In our single site model this is achieved in the simplest way. 

The above analogy between ordinary equilibrium phase transitions and
transitions in the large deviation function of current fluctuations
suggests that the ordinary free energy can be considered as the large
deviation function of density fluctuations, which is indeed the case.
Vice versa, the large deviation of current fluctuations $\hat{e}(j)$
can be interpreted as some kind of {}``dynamical free energy''.
The main difference between the two cases is that the pressure
can easily be controlled and measured in an experiment, whereas that
is not the case with $\lambda$. So far it seems that the only way
to make measurements for $e(\lambda)$ or $\hat{e}(j)$ is the highly inefficient method
of measuring the probabilities of large current fluctuations in a system where $\lambda=0$
(zero pressure). However in the next section, we will see that with
the help of a neat trick $\lambda$ becomes adjustable in computer
simulations. Still, the question of the (experimental) physical
interpretation of this parameter remains open.

\subsection{Range of validity of GC symmetry}

Armed with the detailed knowledge of the phase diagram for our single-site model,
we now return to the original question of the validity of the fluctuation
theorem.  For the GC symmetry to hold for currents of magnitude $j$, we require that both $j$ and $-j$
are in phases B or C (in which $e(\lambda)$ is given by the lowest
eigenvalue of $\tilde{H}$).   This immediately implies that the
symmetry is seen only in the restricted interval
$[-j_\mathrm{max},j_\mathrm{max}]$ where
\begin{equation} \label{e:GCrange}
j_\mathrm{max} \equiv \min (j_b,-j_e).
\end{equation}
This is shown as the shaded regime in figure~\ref{fig:mu-j}.  Note
that $j_\mathrm{max}$ depends on the initial distribution and that the GC
symmetry is not seen at all above some critical value of $x$. 

Now let us investigate the ratio of probabilities for forward and
backward currents outside the symmetry regime.  For non-zero $x$, then one sees a crossover to
\begin{eqnarray}
\fl \hat{e}(-j)-\hat{e}(j)&=f_{-j}(\alpha,\gamma)+\beta(1-x)+\delta(1-x^{-1})
- j \ln x - f_j(\alpha,\gamma) \cr
&=\beta(1-x)+\delta(1-x^{-1}) + j \ln \left( \frac{\alpha}{\gamma x} \right).
\end{eqnarray}
In other words, for large $j$, the quantity $\hat{e}(-j)-\hat{e}(j)$ is linear with slope 
$\ln \left( \frac{\alpha}{\gamma x} \right)$.  At first glance, this
may appear to be the extended fluctuation theorem of
section~\ref{ss:EFT}.  However, the slope of the linear section is not
as predicted there---in particular, it is not zero for an initial
steady-state distribution (i.e., the ratio of forward and backward
currents does not saturate to a constant value).  We emphasize also that $\hat{e}(j)$ itself does
not become linear for large $|j|$ as it does in the case of
boundary terms which are independent of the bulk contribution.  It
would be interesting to develop a general argument to predict the
behaviour in cases, such as this, where correlations are important.

\subsection{Generalizations}

\label{ss:gen}

Although the calculations of appendix~\ref{appendix} and section~\ref{ss:phase}
were for a single-site model with the rates $w_{n}=1$ and a Boltzmann
initial distribution~(\ref{e:boltz}) we argue here that they also
have implications for more general single-site models.

Firstly, we note that results for the case $w_{n}=a$ where $a$ is
any finite positive constant are trivially given by rescaling $\beta\to a\beta$
and $\gamma\to a\gamma$. More generally, we argue that for the long-time
behaviour only the large-$n$ asymptotics of $w_{n}$ are relevant
and thus the results for any bounded $w_{n}$ function $\lim_{n\to\infty}w_{n}=a$
are obtained by the same rescaling of $\beta$ and $\gamma$. To see
this note that the lowest eigenvalue in the gapped state~(\ref{e:elamb}) is independent
of $w_{n}$ and the condition for occurrence of instantaneous condensates
(divergence of the scalar products) depends only on the behaviour
for $n\to\infty$ as does the asymptotic current distribution out
from such an instantaneous condensate. However the convergence to the long-time
limiting behaviour is expected to depend on the form of $w_{n}$ making
direct comparison of finite-time simulation results difficult. 

By a similar argument, we expect any initial distribution with the
same large-$n$ behaviour to lead to the same current large deviations.
Since the elements of the lowest eigenvector $|\psi_0\rangle$ fall off exponentially with $n$,
this means that any initial distribution with super-exponential decay should give
the same result as the empty initial site (or any other fixed initial
configuration), i.e., $x=0$. Similarly, if the weight of configurations in the initial state decays slower than exponentially, then this corresponds to the case $x=1$.

Finally, we remark that results for the current across the output bond
can always be obtained by the
replacements: $\alpha\leftrightarrow\delta$, $\beta\leftrightarrow\gamma$,
$p\leftrightarrow q$, $\lambda\leftrightarrow-\lambda$, $j\leftrightarrow-j$,
i.e., by left-right reflection.
In section~\ref{s:large} we discuss extensions to larger systems.

\section{Numerical results for $e(\lambda)$}

\label{s:num}

In~\cite{Harris2006}, the analytical prediction for $\hat{e}(j)$ was
checked via direct Monte Carlo simulation. However, it is difficult
to get high quality data for long-times since one is looking for exponentially
unlikely events. In this section we demonstrate instead the application
of a recent algorithm by Giardin\`a, Kurchan and Peliti~\cite{Giardina06}
to calculate $e(\lambda)$ directly.

\subsection{Cloning algorithm}

The method is based on an alternative interpretation of (\ref{eq:Htilde}).
Here $\tilde{H}$ acts as the generator of some kind of time evolution.
However, it cannot be an ordinary stochastic evolution operator since
it does not satisfy the normalization condition (\ref{Hnormalization}),
therefore it does not conserve probability. The idea is to consider
an ensemble of $N$ identical systems (clones), where the number of
{}``individuals'' being in the same configuration $\sigma$ is denoted
by $P_{\sigma}$. Here the size $N$ of the {}``population'' can
vary in time, which means that there is no conservation law for the
vector $|P\rangle$ as it is not a probability vector anymore ($\langle s|P\rangle\neq 1$).
The off-diagonal elements of $\tilde{H}$ give the currents from one
configuration to another. The fact that $\tilde{H}$ is not normalized
means that the sum of the currents from a specific configuration into
any other is not necessarily equal to the loss in that configuration.
This means that individuals can reproduce (i.e. introduce another
copy of the system prepared in the same configuration as the ancestor
by increasing $N$ by one) or die at given rates, where this rate
depends on the configuration $\sigma$, and is given by $\sum_{\sigma'}\tilde{H}_{\sigma'\sigma}$. 

The process described above can easily be simulated by Monte Carlo
methods. The quantity to measure is the average rate of growth of
the whole population for large times, which directly gives $e(\lambda)$.
In practice however, one chooses a sufficiently large $N$ which is
kept constant by renormalizing the size of the population after each
``birth'' or ``death'' (while keeping a record of growth).
The method was originally introduced for discrete-time update (for
details see \cite{Giardina06}) but is straightforwardly modified to
the continuous time case (see also \cite{Lecomte07}).

Note that $\tilde{H}$ does not give the full probabilistic description
of the above-defined stochastic cloning process. The deterministic time evolution
(often called rate equation) given by $\tilde{H}$ refers only to
the averages in a large population ($N\to\infty$). Therefore, in order to obtain reliable results one has to reduce the possible fluctuations of the (cloning) process by choosing large $N$. 

In a modified version of the algorithm the mean (integrated) current
is measured instead of the growth rate. This gives the derivative
of $e(\lambda)$, which can be then numerically integrated (with the
fixed condition
$e(0)=0$). We note that, in this algorithm, descendants of a given
clone inherit not only the configuration of the parent but also the
actual value of the parent's integrated current. The advantage of this
modified algorithm is that it gives less noise.

\subsection{Results for the one-site ZRP with empty initial condition}

\begin{figure}
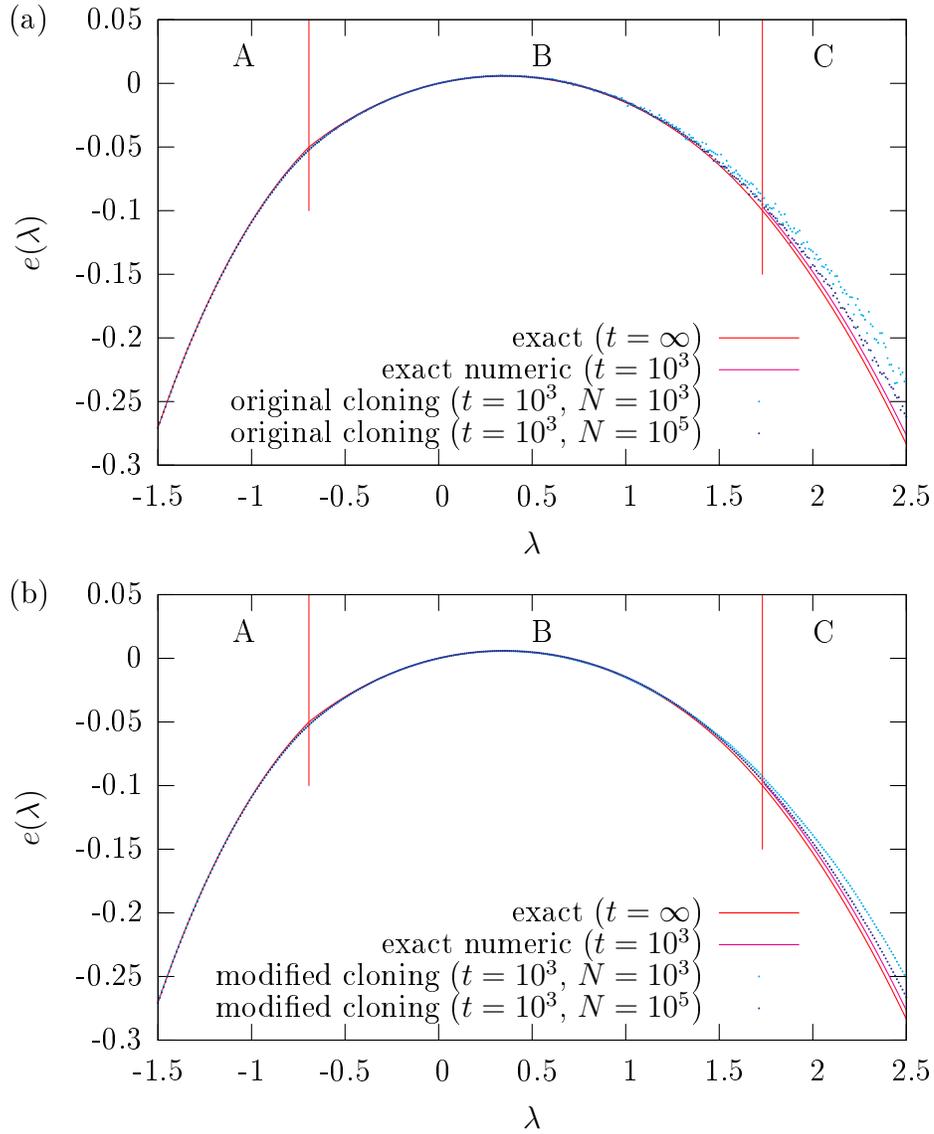

\begin{center}
\input{figure3_orig.tex}
\input{figure3_mod.tex}
\end{center}
\caption{\label{fig:figure3} Cloning simulation results for the one-site
ZRP with empty initial condition compared to the exact analytical
results. The plots show the result of two versions of the algorithm,
both with two different values of $N$. The original cloning algorithm
(a) produces more noise in the gapless regime than a variant
of it (b) whereby the derivative of the function $e(\lambda)$
is measured and then numerically integrated to get $e(\lambda)$.
In the gapless case the convergence in $N$ is slow. For reasonable
values of $N$ there is a systematic deviation (overestimation) from
the exact result. Parameters: $\alpha=\gamma=\delta=0.1$,
$\beta=0.2$.}
\end{figure}

We performed the above programme for the one-site ZRP. Since $e(\lambda)$
is known exactly, this should be considered as a test of the cloning
method. Results are shown in figure \ref{fig:figure3}(a) for the case
of an empty initial site. It can be seen that the resulting data points
for $N=10^{3}$ and $t=10^{3}$ lie very close to the exact ($t=\infty$)
results in phases A and B. However, in phase C there is a significant
deviation and the difference decreases with increasing
$N$. A systematic analysis of the $N$-dependence of the cloning method is still lacking so it is not obvious how to set the value of $N$ in general to provide a good balance between accuracy and simulation speed. 

Our numerical results suggest that in the gapless phase
the simulation could be very sensitive to the value of $N$
and could deliver unreliable results for the accessible range. We
believe that this limitation of the cloning method is related to the
gapless spectrum and would not show up in cases where the state space
is finite. 

Figure \ref{fig:figure3}(b) shows the results of a modified cloning algorithm, where the derivative of $e(\lambda)$ was measured directly and then numerically integrated. Despite the smoother results the same type of discrepancy shows up in this version of the simulation. 

\begin{figure}
\begin{center}
\input{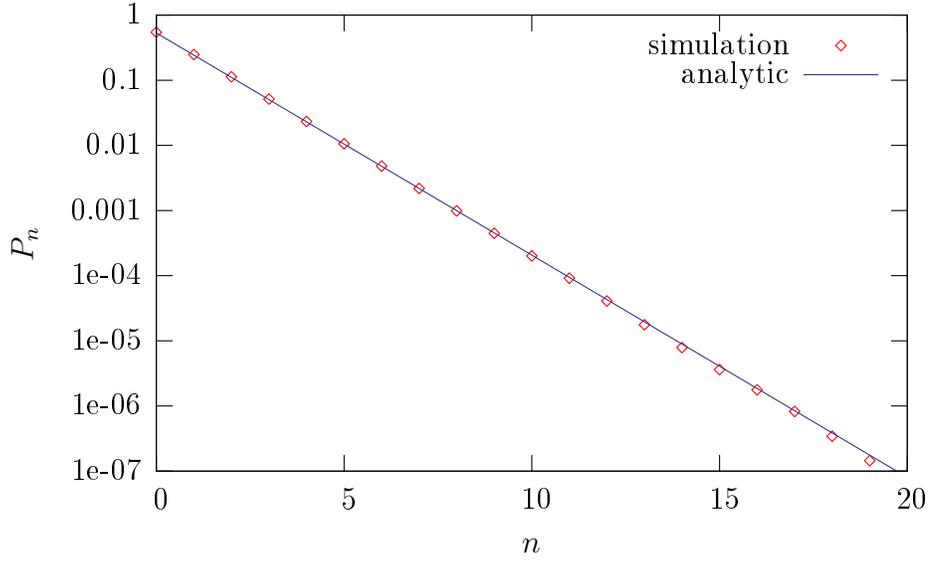}
\end{center}
\caption{Distribution of the occupation number in the {}``steady state''
of the cloning algorithm at $\lambda=1.0$ (phase B) compared to the
analytical results (ground-state of $\tilde{H}$). Parameters: $\alpha=\gamma=\delta=0.1$, $\beta=0.2$, $t=10000$, $N=10^{5}$. \label{fig:distribution1}}
\end{figure}

\begin{figure}
\begin{center}
\input{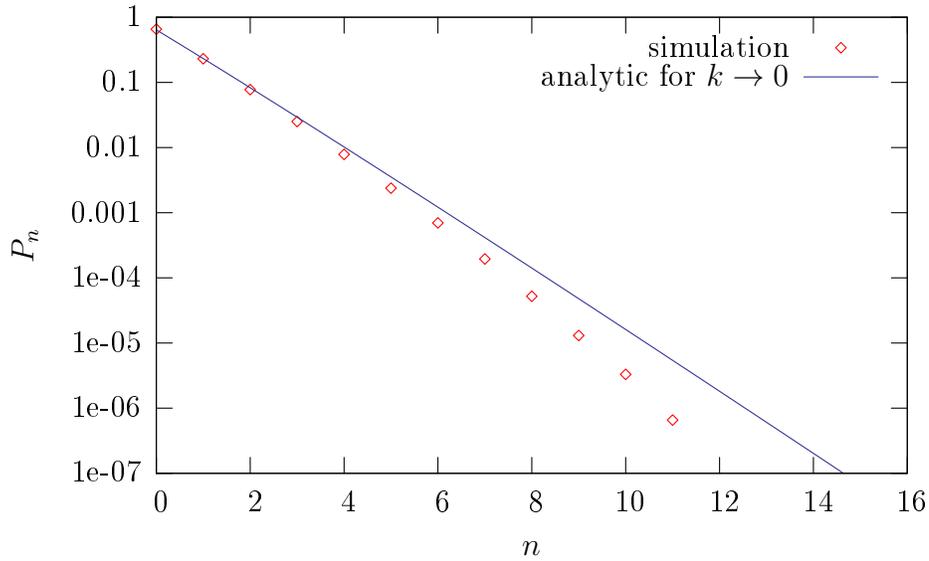}
\end{center}
\caption{Distribution of the occupation number in the ``steady state''
of the cloning algorithm at $\lambda=2.2$ (phase C) compared to the
analytical results. Taking the limit of expression (\ref{eq:varphi})
at $k\to0$ together with (\ref{eq:transform}) one obtains $\langle n|\psi_{k\to0}\rangle\sim\phi^{-n}(n+y/(y-1))$.
The algorithm seems to fail in this gapless region and the distribution
does not converge to the ground-state of $\tilde{H}(\lambda)$. Parameters:
$\alpha=\gamma=\delta=0.1$, $\beta=0.2$, $t=10000$, $N=10^{5}$.  \label{fig:distribution2}}
\end{figure}

In the stationary state of the cloning process (if it exists), the distribution of the occupation number of clones is expected to follow the ground-state of $\tilde{H}$. We measured this distribution in a representative point of phase B ($\lambda=1.0$, $x=0$) and C ($\lambda=2.2$, $x=0$). Results are shown in figure \ref{fig:distribution1}-\ref{fig:distribution2}. The measured distribution in phase B follows the one suggested by the theory. In phase C (gapless phase) however, there is a significant difference between the theory and the measurement. This suggests that in the gapless phase the algorithm fails to find the true ground-state, hence $e(\lambda)$ is systematically overestimated in the measurement. This is consistent with the results shown in figure \ref{fig:figure3}. We note that in phases A and D there is no steady state of the cloning process.

As discussed above, the cloning method strongly relies on the assumption that the number of clones is infinite. It is highly non-trivial to determine what type of correction appears if $N$ is finite (as in computer simulations). Our numerical results show that this correction is much higher in the gapless phase. This is possibly due to the fact that with a finite number of clones one cannot recover the exact distribution but some fluctuations are introduced. In the case of a gapped spectrum the system has a strong tendency to the ground state, therefore these fluctuations are suppressed and do not play a crutial role as long as they are sufficiently small ($N$ is sufficiently large). In the gapless case however, the relaxation to the true ground-state becomes slow and therefore, due to the fluctuations, states other than the ground state are also represented in the sample with a non-negligible weight (see figures \ref{fig:distribution1}-\ref{fig:distribution2}).

\begin{figure}
\begin{center}
\input{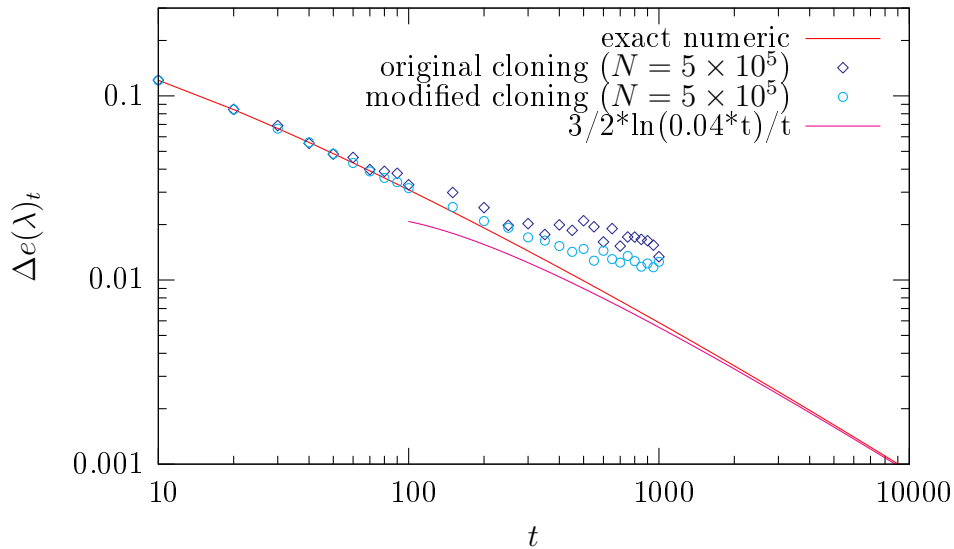}
\end{center}
\caption{\label{fig:finite_time} Plotted is the finite-time correction $\Delta e(\lambda)_{t}=e(\lambda)_{t}-e(\lambda)$
against $t$ in the gapless phase at $\lambda=2.2$. Exact numerical
calculations are compared to the simulation results. The cloning method
(with fixed $N$) becomes unreliable for large $t$. Parameters: $\alpha=\gamma=\delta=0.1$,
$\beta=0.2$.}
\end{figure}

It is also possible to define $e(\lambda)$ for finite time, which we denote
by $e(\lambda)_{t}$. 
\begin{equation}
e(\lambda)_{t}=-\frac{1}{t}\ln\langle \rme^{-\lambda J}\rangle
\end{equation}
Since measurements can be made only for finite times, the corrections
to $e(\lambda)$ are of great interest. It is relatively easy to calculate
the finite-time corrections in the single-site model based on (\ref{eq:final}).
In phases A-B-D the leading contribution comes from a pole which
leads to an ${\cal O}(1/t)$ correction in $e(\lambda)_{t}$. In the
gapless phase C the leading contribution is given by the saddle
point integration. Since the first order term vanishes here the dominant
contribution gives ${\cal O}(t^{-2/3}\exp(-e(\lambda)t))$ in (\ref{eq:final}),
which leads to a $(2/3)\ln(Ct)/t$ correction, where $C$ is some constant.
 We performed exact numeric
calculations of $e(\lambda)_{t}$ by evaluating the integrals in (\ref{eq:final})
and this is in good agreement with our analytical findings. For details
see figure \ref{fig:finite_time}.

\subsection{Non-empty initial condition}

It is, in principle, possible to start the system from any initial condition
in the cloning algorithm. Ideally, one would hope that the method
reproduces the exact results in this regime. However, since the number
of clones is finite, the initial distribution has always a finite
(although arbitrarily large) cutoff. Although this is a minor
change in the initial distribution, it becomes crucially important
when measuring large deviations. A long time measurement for an initial
distribution with a finite cutoff gives the same results as for an
initially empty lattice. For an initial distribution with exponential
tails (the case that we consider) one would need $N\sim\exp(t)$ number
of clones in a measurement of length $t$. This is practically unreachable
for reasonably large $t$. For this reason it is not surprising that
the algorithm breaks down in the initial-state-dependent phase.

\section{Larger systems}
\label{s:large}

In this section we extend our discussion to larger systems in order to
demonstrate the generality of the fluctuation theorem
 breakdown for the
current fluctuations.  We consider
the $L$-site zero-range process with parameters as defined in
section~\ref{s:ZRP} and take again $w_n=1$ (but expect qualitatively the
same results for any bounded $w_n$).  For definiteness we assume that
the boundary parameters have been chosen to give
a well-defined steady
state with mean (``forward'') current to the right.  Once again, we
focus on the behaviour of current fluctuations across the input
(``0th'') bond but indicate also how to extend our approach to treat
the fluctuations across bulk bonds.  In section~\ref{ss:Lsite}, we first
outline our general method, in particular the use of a powerful
mapping to effective one-site systems.  This leads to a statement about
the regime of validity of the GC symmetry (section~\ref{ss:GCvalid}) and some comments on
generalizations (section~\ref{ss:generalizations}).  Finally, in section~\ref{ss:2site}, we give
explicit results for a two-site system and make comparisons with
numerical data.

\subsection{General approach}
\label{ss:Lsite}

We remind the reader that, for a ZRP of arbitrary size, one can
rigorously calculate the lowest eigenvalue $\epsilon_0$ \emph{in the
gapped phase} of the modified Hamiltonian measuring the input current
together with the associated left and right eigenvectors $\langle
\psi_0 |$ and $| \psi_0 \rangle$.  Details of the calculations are
given in~\cite{Me05}, pertinent results are summarized in
section~\ref{s:ZRP} above.  In order to determine the limit of the GC
symmetry one has to ascertain the values of $\lambda$ at which the
scalar products $\langle s | \psi_0 \rangle$ and $\langle \psi_0 | P_0
\rangle $ diverge whilst keeping in mind that the divergence of the
normalization of the gapped eigenvalue $\langle \psi_0 | \psi_0
\rangle$ indicates the cross-over to a gapless regime.  To determine
the behaviour of current fluctuations when these scalar products diverge
we can appeal to the heuristic argument based on instantaneous
condensates.

This process is considerably simplified by relating the current
fluctuations in an $L$-site system to those in a one-site system with
effective parameters. We note that with the definitions
\begin{eqnarray}
\alpha_l&= \frac{\alpha(p/q)^{l-1} (p-q)}{(p-q+\gamma)(p/q)^{l-1}-\gamma} \label{e:alphap} \\
\beta_l &= \frac{\beta(p/q)^{L-l} (p-q)}{p-q-\beta+\beta(p/q)^{L-l}} \\
\gamma_l&= \frac{\gamma  (p-q)}{(p-q+\gamma)(p/q)^{l-1}-\gamma} \\
\delta_l &= \frac{\delta (p-q)}{p-q-\beta+\beta(p/q)^{L-l}}, \label{e:deltap}
\end{eqnarray}
the lowest eigenvalue~(\ref{e:elamb}) in the gapped phase of the
modified Hamiltonian for an $L$-site system~(\ref{e:H0}) can be
written in the form
\begin{equation} \label{e:gap}
\epsilon_0
=\frac{\alpha_l\beta_l}{\beta_l+\gamma_l}(1-\rme^{-\lambda})+\frac{\gamma_l\delta_l}{\beta_l+\gamma_l}(1-\rme^\lambda),
\end{equation}
and the corresponding left and right eigenvectors are defined by fugacities
\begin{eqnarray}
z_l&=\frac{\alpha_l \rme^{-\lambda} + \delta_l}{\beta_l+\gamma_l} \label{e:z1}\\
\tilde{z}_l&=\frac{\beta_l + \gamma_l \rme^\lambda}{\beta_l+\gamma_l}. \label{e:z2}
\end{eqnarray}
In other words, the one-site marginal for site $l$ and the eigenvalue
have exactly the same form as in the one site problem with effective
$\alpha_l$ and $\gamma_l$ ($\beta_l$ and $\delta_l$) determined by all
the rates to the left (right) of $l$.  This fact relies on the
product-state character of the lowest eigenvector of the modified
Hamiltonian corresponding to the current being measured.
Notice that, by construction,
$\alpha_1=\alpha$, $\gamma_L=\gamma$, $\beta_L = \beta$ and
$\delta_L=\delta$. We also remark that the parameter combinations ${\alpha_l\beta_l}/(\beta_l+\gamma_l)$ and ${\gamma_l\delta_l}/(\beta_l+\gamma_l)$ are independent of $l$ and are equal to the left/right hopping rates of an $N$ particle bound state in the corresponding exclusion process. Equivalently, in current space we have
\begin{equation}
\hat{e}(j)=  f_j\!\!\left(\frac{\alpha_l\beta_l}{\beta_l+\gamma_l},\frac{\gamma_l\delta_l}{\beta_l+\gamma_l}\right). 
\end{equation}
 This important observation allows us to utilize the exact results
already obtained for the one-site case to build up the phase diagram
for an $L$-site system.\footnote{Although our results
  can be applied to large systems by taking the thermodynamic limit
  $L\to\infty$, this limit does not necessarily commute with the
  long-time limit $t \to \infty$.  Our analysis therefore does not
  address the form of current fluctuations in a genuinely infinite
  system (taking the limit $L \to \infty$ first, followed by $t \to \infty$).}

Recall that $\lambda=0$ corresponds to the mean current.  The asymptotic
probability of seeing a current fluctuation larger than the mean is
given by the behaviour of $e(\lambda)$ for $\lambda$ negative.   To
determine the $\lambda<0$ part of the phase diagram we carry out the
following procedure:

\begin{enumerate}
\item Start with $\lambda=0$ and $e(\lambda)$ given by $\epsilon_0$
  of~(\ref{e:gap}).
\item Decrease $\lambda$ until $\langle s | \psi_0 \rangle$ diverges
  on one of the sites, which we label $l_1$.  From the one-site picture we immediately see
  that this will happen at
\begin{equation}
\lambda_{1}(l_1) \equiv \lambda_1(\alpha_{l_1},\beta_{l_1},\gamma_{l_1},\delta_{l_1})
\end{equation}
with $ \lambda_1$ defined as in Table~\ref{t:lam}.  Physically, we
argue (just as in the single-site case) that this divergence
corresponds to the ``piling-up'' of particles on site $l_1$.  For
$\lambda<\lambda_{1}(l_1)$ the current fluctuations (across the input
bond) only depend on the part of the system to the left of site $l_1$
 and we thus expect a crossover to
\begin{equation}
e(\lambda)=\alpha_{l_1}(1-\rme^{-\lambda})+\gamma_{l_1}(1-\rme^\lambda),
\end{equation}
which corresponds to a current large deviation function.
\begin{equation}
\hat{e}(j)=  f_j(\alpha_{l_1},\gamma_{l_1}).
\end{equation}
Just as in the single-site case, $e_0(\lambda)$ is continuous but not
differentiable at $\lambda_{1}(l_1)$ so the two phases in current space
are separated by a linear transition regime whose explicit form is
simply obtained from the effective single-site picture.
\label{step:2}
\item Now, with such an ``instantaneous condensate'' on site $l_1$, the
  left-hand part of the system looks just like a system of size $l_1-1$ with right-hand boundary rates $p$
  and $q$.  We can then write down the new ground state $| \psi_0 \rangle$ in
  terms of redefined effective parameters $\alpha_l$,
  $\beta_l$, $\gamma_l$ and $\delta_l$. \label{step:3}
\item Repeat steps~(\ref{step:2}) and~(\ref{step:3}) recursively until...
\item At some value of $\lambda$,  $\langle s | \psi_0 \rangle$ (with $|\psi_0\rangle$ a
  function of the effective parameters) diverges on site 1 and
  the large deviation function takes the form
\begin{eqnarray}
e(\lambda)&=\alpha(1-\rme^{-\lambda})+\gamma(1-\rme^\lambda) \\
\hat{e}(j)&=  f_j(\alpha,\gamma).
\end{eqnarray}
(again with a linear transition regime in $j$-space).  In other words, for very large forward currents, only the hopping parameters across the input bond are significant.
\end{enumerate}

In a similar fashion one can investigate what happens for currents smaller
than the mean by increasing $\lambda$ from zero. For simplicity, let us first take
a fixed initial configuration so that $\langle \psi_0 |
P_0 \rangle $ is always finite.  At
$\lambda_{2}(l) \equiv \lambda_2(\alpha_l,\beta_l,\gamma_l,\delta_l)$,
$\langle \psi_0 | \psi_0 \rangle $ diverges on site $l$, corresponding to a
transition to a gapless phase.  The minimum value of $\lambda_{2}(l)$
occurs at some $l_2$ (which depends on the parameters of the model) so we first see a crossover to
\begin{eqnarray}
e(\lambda)&=\alpha_{l_2}+\beta_{l_2}+\gamma_{l_2}+\delta_{l_2}-2\sqrt{(\alpha_{l_2} \rme^{-\lambda} +\delta_{l_2}) (\beta_{l_2} + \gamma_{l_2} \rme^\lambda)}\\
\hat{e}(j)&=f_j(\alpha_{l_2},\gamma_{l_2})+f_j(\beta_{l_2},\delta_{l_2}). 
\end{eqnarray}
In other words, to sustain a large backwards current we have an instantaneous
condensate on site $l_2$ and the product of current distributions
across the two independent parts of the system.  For increasing
backwards currents, one can then repeat the procedure for the two
subsystems (each with redefined effective parameters), increasing
$\lambda$ and looking for the next site where $\langle \psi_0 | \psi_0
\rangle $ diverges.
Eventually, for very large backward currents we expect instantaneous
condensates on all sites and a current large deviation function given
by
\begin{equation}
\hat{e}(j) =f_j(\alpha,\gamma)+(L-1)f_j(p,q)+f_j(\beta,\delta)
\end{equation}

For a distribution over initial configurations, the situation is more
complicated but again we can make progress based on the effective
one-site picture.  For an initial particle distribution which is a
product measure of Boltzmann distributions with site dependent
fugacity $x_l$, then $\langle \psi_0 | P_0 \rangle $ will diverge on
the $l$th site at $\lambda_{4}(l) \equiv \lambda_4(\alpha_l,\beta_l,\gamma_l,\delta_l,x_l)$
leading to a crossover to an initial-state dependent regime. If $l_4$
is the (initial-condition-dependent) value of $l$ corresponding to the
minimum of $\lambda_{4}(l)$ then, for large initial fugacities we expect to find first a transition
to
\begin{eqnarray} \fl
e(\lambda)&=\alpha_{l_4}+\beta_{l_4}+\gamma_{l_4}+\delta_{l_4}-(\beta_{l_4} + \gamma_{l_4} \rme^\lambda) x_{l_4}  - (\alpha_{l_4} \rme^{-\lambda} +\delta_{l_4} ) x_{l_4} ^{-1}\\
\fl
\hat{e}(j)&=f_j(\alpha_{l_4} ,\gamma_{l_4} )+\beta_{l_4} (1- x_{l_4}
)+\delta_{l_4} (1- x_{l_4}^{-1}) + j \ln x_{l_4}. 
\end{eqnarray}
Once again, we see a factorization of the asymptotic current
distribution across the two sub-systems to the left and right of
$l_4$.  In principal, one can try to build up the complete phase
diagram by next checking for the divergence of $\langle \psi_0 | P_0
\rangle $ on other sites as $\lambda$ is further increased.  Another
scenario, for small initial fugacities, involves first a transition to
a gapless phase and then a transition to the initial condition
dependent phase when $\lambda=\lambda_{3}(l)$ as a function of the
redefined effective rates.

\subsection{Range of validity of GC symmetry}
\label{ss:GCvalid}

The effective one-site picture provides an elegant way to summarize
the range of validity for the GC symmetry in ZRPs of arbitrary size.
By comparison with (\ref{e:GCrange}) and with the shaded regime in figure~\ref{fig:mu-j}, we see that, for given initial fugacities $\{x\}$, the symmetry
relation for input current will only be obeyed in the following
current range:
\begin{equation}
|j| \leq \min
 (j_b(\alpha_{l_1},\beta_{l_1},\gamma_{l_1},\delta_{l_1}),-j_e(\alpha_{l_4},\beta_{l_4},\gamma_{l_4},\delta_{l_4},x_{l_4})) \label{e:gccon}
 \end{equation}
 where $l_i$ is the value of $l$ at which $|\lambda_{i}(l)|$ takes its
minimum value as a function of the effective rates given
in~(\ref{e:alphap})--(\ref{e:deltap}).  In the case where $j_e$ is
positive for some $l$ then the symmetry will not be observed at
all.

\subsection{Further generalizations}
\label{ss:generalizations}

We remark that the same condition as (\ref{e:gccon}) should also
apply for ZRPs with quenched disorder (i.e., with $p$ and $q$ bond dependent)---one needs only to determine the appropriate form for
the effective rates.  Indeed, building on the recent determination of the stationary
state for such systems~\cite{Pulkkinen2007}, we can show that the relevant
parameter combinations are 
\begin{eqnarray}
\label{alphal}
\alpha_l&= \frac{\prod_{i=0}^{l-1} \frac{p_i}{q_i}}{\sum_{i=0}^{l-1}
  \frac{1}{q_{l-i-1}}\prod_{k=1}^{i} \frac{p_{l-k}}{q_{l-k}}}   \\
\label{betal}
\beta_l &= \frac{1}{\sum_{i=0}^{L-l}\frac{1}{p_{l+i}}\prod_{k=0}^{i-1}
  \frac{q_{l+k}}{p_{l+k}}}  \\
\label{gammal}
\gamma_l &= \frac{1}{\sum_{i=0}^{l-1}\frac{1}{q_{l-i-1}}\prod_{k=1}^{i} \frac{p_{l-k}}{q_{l-k}}}  \\
\label{deltal}
\delta_l &= \frac{\prod_{i=0}^{L-l} \frac{q_{l+i}}{p_{l+i}}}{\sum_{i=0}^{L-l}\frac{1}{p_{l+i}}\prod_{k=0}^{i-1} \frac{q_{l+k}}{p_{l+k}}} 
\end{eqnarray}
where $p_l$ ($q_l$) is the right (left) hopping rate across bond $l$
and by definition $p_0=\alpha$, $q_0=\gamma$, $p_L=\beta$ and $q_L=\delta$.
One can state (\ref{alphal}-\ref{deltal}) alternatively as a simple recursion relation:
\begin{eqnarray}
\alpha_{l+1}&=\frac{\alpha_l p_l}{\gamma_l + p_l} \\
\beta_{l-1}&=\frac{\beta_l p_{l-1}}{\beta_l + q_{l-1}} \\
\gamma_{l+1}&=\frac{\gamma_l q_l}{\gamma_l + p_l} \\
\delta_{l-1}&=\frac{\delta_l q_{l-1}}{\beta_l + q_{l-1}}
\end{eqnarray}
Not surprisingly, these are again the renormalized two-particle
hopping rates. 

We conclude this subsection by indicating how to extend our approach to treat the
current fluctuations across other bonds.  In this case, the effective
parameters are unchanged, but in all sites to the left of the bond in
question the fugacities~(\ref{e:z1}) and~(\ref{e:z2}) must be replaced
by
\begin{eqnarray}
z_l&=\frac{\alpha_l + \delta_l \rme^\lambda}{\beta_l+\gamma_l} \\
\tilde{z}_l&=\frac{\beta_l + \gamma_l \rme^{-\lambda}}{\beta_l+\gamma_l}. 
\end{eqnarray}
corresponding to the expressions for the \emph{output} bond of a
single-site system.

\subsection{Example: Two-site system}
\label{ss:2site}

In principle, the general approach outlined above can be used to
determine the behaviour of current fluctuations in ZRPs of
arbitrary size.  However the implementation rapidly becomes tedious and
the phase diagram complicated.  Here, as a simple test case, we present
results for a two-site system with empty initial condition ($x_l=0$ for all $l$).

It is obvious that for any choice of parameters obeying $\alpha-\gamma < p - q < \beta -
\delta$, the system has a well-defined steady state (i.e., no boundary condensation).
One also gets a stationary state if the conditions
\begin{equation}
\alpha-\gamma < \beta-\delta < p-q \label{e:conssa}
\end{equation}
and
\begin{equation}
\frac{\alpha p}{p+\gamma}-\frac{\gamma q}{p+\gamma}<\beta-\gamma \label{e:conss}
\end{equation}
are obeyed.  These inequalities can easily be understood in the exclusion picture where
the first two particles form a bound state with hopping rates
$\frac{\alpha p}{p+\gamma}$ and $\frac{\gamma q}{p+\gamma}$. The
condition for the bound state of this and the remaining single
particle is just~(\ref{e:conss}).

For definiteness, in the remainder of the discussion we assume that
the rates obey~(\ref{e:conssa}) and~(\ref{e:conss}).  In this case, as $\lambda$ is decreased from 0, $\langle s | \psi_0  \rangle$
diverges first on site 2 (i.e., $l_1=2$).  Similarly, as $\lambda$ is
increased from 0, $\langle \psi_0 | \psi_0 \rangle $ also diverges first on site
2 (i.e., $l_2=2$).  Appealing to the heuristic ``instantaneous
condensates'' picture we then obtain the following different regimes
for the current large deviation function.
\begin{equation} 
\hat{e}(j)= 
\left\lbrace \begin{array}{ll}
f_j(\alpha,\gamma) & 
\mathrm{A2} \\
f_j\!\!\left(\frac{\alpha p}{p+\gamma},\frac{\gamma q}{p+\gamma}\right) &  
\mathrm{A1} \\
f_j\!\!\left(\frac{\alpha\beta p}{\gamma q + \beta p + \beta \gamma},\frac{\gamma\delta q}{\gamma q + \beta p + \beta \gamma}\right) & 
\mathrm{B}\\
f_j\!\!\left(\frac{\alpha p}{p+\gamma},\frac{\gamma q}{p +
  \gamma}\right) + f_j(\beta,\delta) & \mathrm{C1} \\
f_j(\alpha,\gamma)+f_j(p,q)+ f_j(\beta,\delta) & 
\mathrm{C2}\\
\end{array}\right.,  \label{e:ejres2} 
\end{equation}
with
\begin{equation}
\begin{array}{rl}
 \mathrm{A2:}	& j_a(\alpha,p,\gamma,q)<j \\
 \mathrm{A1:}	& j_a\left(\frac{\alpha
  p}{p+\gamma},\beta,\frac{\gamma q}{p +\gamma},\delta\right)
<j<j_b(\alpha,p,\gamma,q) \\
 \mathrm{B:}	& j_c\left(\frac{\alpha p}{p+\gamma},\beta,\frac{\gamma q}{p +\gamma},\delta\right)<j<j_b\left(\frac{\alpha p}{p+\gamma},\beta,\frac{\gamma q}{p +\gamma},\delta\right) \\
 \mathrm{C1:}	& j_c(\alpha,p,\gamma,q)<j<j_c\left(\frac{\alpha
  p}{p+\gamma},\beta,\frac{\gamma q}{p +\gamma},\delta\right) \\
 \mathrm{C2:}	& j>j_c(\alpha,p,\gamma,q)
\end{array}.
\end{equation} 
Here the currents $j_a$, $j_b$ and $j_c$ are as defined in
Table~\ref{t:lam}; we label the phases in analogy to A, B and C in the
single-site case and note that there are intermediate transition
regions at the A2--A1 and A1--B crossovers.  From the argument in~\ref{ss:GCvalid} it is clear that the Gallavotti-Cohen symmetry should
hold for small currents $j<j_b\left(\frac{\alpha p}{p+\gamma},\beta,\frac{\gamma q}{p
+\gamma},\delta,\right)$.

\begin{figure}
\begin{center}
\input{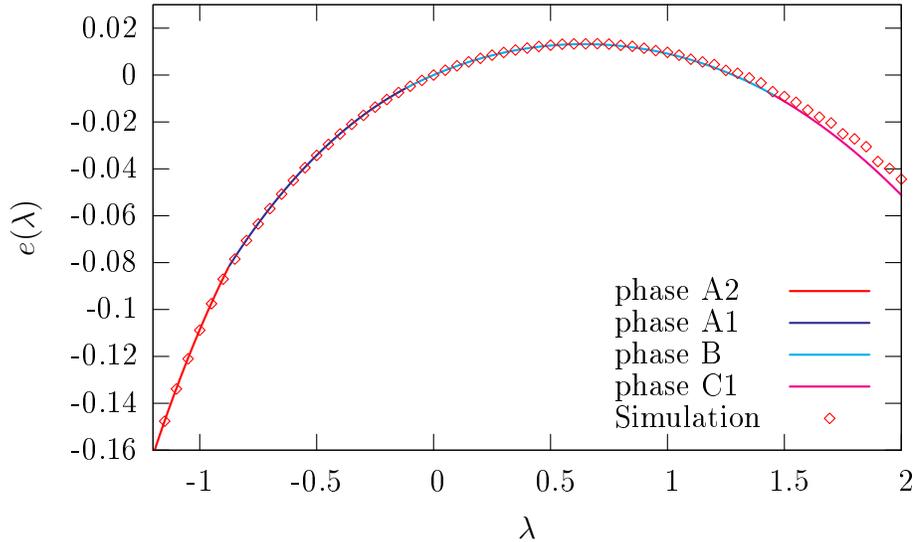}
\end{center}
\caption{Cloning simulation results for two-site ZRP with empty
  initial condition compared with exact analytical results.  Agreement is excellent except
  in the gapless phase.
  Parameters: $\alpha=\gamma=\delta=q=0.1$, $\beta=0.15$, $p=0.24$, $t=5000$, $N=10^4$.  \label{f:2sitecf}}
\end{figure}

Finally, we compare the Legendre transform of~(\ref{e:ejres2}) with
results for $e(\lambda)$ obtained via the cloning algorithm (see
section~\ref{s:num}).  As shown in figure~\ref{f:2sitecf}, 
we find excellent
agreement except for $\lambda > \lambda_2 \left(\frac{\alpha
  p}{p+\gamma},\beta,\frac{\gamma q}{p +\gamma},\delta\right)$ where
$e(\lambda)$ is given by the lowest eigenvalue of a gapless
spectrum.  Just as in the single-site case we attribute
this to limitations imposed by the finite number of clones.
 
\section{Summary}

\label{s:conc}

This paper focuses on the Gallavotti-Cohen (GC) fluctuation symmetry in the
context of continuous-time Markov processes. In particular, it offers
a contribution towards understanding the potential breakdown of this
symmetry in systems with \emph{infinite state space}.   To set the scene for
this, we first discussed how the symmetry is manifested for the large
deviation function of currents in systems with \emph{finite state space}. Our
proof was based on that of Lebowitz and Spohn \cite{Lebowitz99} (or
Derrida et al.~\cite{Derrida2004}) but slightly extends their work by
considering more general currents. In the notational framework of this
paper, the action functional of Lebowitz and Spohn can be obtained by
choosing
$\Theta_{\sigma,\sigma'}=\ln(w_{\sigma',\sigma}/w_{\sigma,\sigma'})$
and the GC symmetry then holds for $J$ with $E=1$. In this case, as
shown in \cite{Lebowitz99}, $J$ can be interpreted as an entropy current.

The above-mentionned proof of the fluctuation relation holds only for systems with finite state
space---in systems with infinite state space the GC symmetry can be
broken\footnote{We remark that an infinite state space is {\em not} a sufficient condition for a breakdown of the Fluctuation Theorem. A counterexample is the ZRP with unbounded $w_n$ as explained in section~\ref{ss:stationary_state}}.  In the spirit of van Zon and Cohen's earlier work for Langevin
systems~\cite{VanZon04,VanZon03}, we next recapitulated the ideas leading
to an ``extended fluctuation theorem'' in which (for stationary
initial state) the ratio of probabilities of given forward ($j$) and
backward ($-j$) currents approaches a constant for large values of
$j$.   We argued that this particular form of breakdown relies on the
independence of bulk and boundary contributions to the current.  The
central aim of this paper was to study analytically a simple model
where this condition is not met---the partially asymmetric Zero Range
Process (ZRP) on an open lattice~\cite{Levine04c,Me05}.

Specifically, we expanded on earlier results in~\cite{Harris2006}, and
showed in detail how to calculate the large deviations of the average
particle current $j$ in the one-site ZRP.  The phase
behaviour of the large deviation function can be physically understood
by considering the possibility for an arbitrarily large number of
particles to pile-up on the site (``instantaneous condensation'').  Even in this single-site case, the phase diagram (parametrized by the time-averaged current and the initial state) shows a complex picture with first and second order phase transitions. These transitions are analogous to ordinary equilibrium phase transitions since formally the current large deviation function can be considered as a kind of dynamical free energy functional. In general, the GC symmetry holds only in a restricted interval $[-j_\mathrm{max},j_\mathrm{max}]$, where $j_\mathrm{max}$ depends on the initial state. The relevant parameter of the initial state, which is denoted by $x$, can be considered as a kind of initial temperature. For large values of $x$ the regime of validity of the fluctuation relation shrinks ($j_\mathrm{max}$ decreases) and reaches zero at a finite $x$. According to the above interpretation, this means that above a critical initial temperature the fluctuation theorem does not hold, even for small currents. 

Although the one-site ZRP might be thought to be
oversimplified, it is very instructive and already exhibits most features of
the multiple-site version, which we have also studied in detail. In
particular, we developed a heuristic argument based on mapping to an
effective one-site system and considering the appearance of
instantaneous condensates.  The phase diagram can, in principle, be
derived for any number of sites but it becomes increasingly
complicated. As an example we showed how the current large deviation
function can be obtained for a two-site ZRP.   Again the phenomenon of
instantaneous condensation is the physical mechanism leading to a
breakdown of the GC fluctuation theorem.

Outside the symmetry regime, we found
that the behaviour is somewhat different to that predicted by the
extended fluctuation theorem of van Zon and Cohen, presumably due to
the correlations (between bulk and boundary terms) in our model.
There are evidently still open questions relating to the
characterization of the fluctuation behaviour of different systems
beyond the limits of Gallavotti-Cohen symmetry.  For example, in
another recent work~\cite{Touchette07b} it was shown that the
distribution of work fluctuations for a Brownian particle with L\'evy
noise (i.e., infinite variance) do not even have a large deviation
form.  It might be interesting to look for analogues of the resulting
``anomalous fluctuation theorem'' in the present framework of many-particle
dynamics described by a master equation.  

Our analytical results also made it possible to test the recently
proposed cloning algorithm~\cite{Giardina06} that measures the
Legendre transform of the large deviation function directly. The
strength of this method is its efficiency and relative simplicity. The
algorithm reproduces our analytical results in several phases.
However, there are notable deviations in a phase where the spectrum of
$\tilde{H}$, the effective Hamiltonian governing the current fluctuations, becomes gapless. Another weakness of the algorithm is that it is unable to capture the initial-state-dependence of the fluctuations. It is only the case of a fixed initial configuration that this method can safely be used. These observations call for the development of a novel numerical algorithm which would not break down in cases where the cloning method does. One possibility here would be to develop a transition path sampling method.

\ack

This work was initiated while R.~J.~H. was working at the
Forschungsentrum J\"ulich; we are very grateful to Gunter Sch\"utz for many helpful discussions.
It is also a pleasure to thank Jorge Kurchan, Vivien Lecomte and
Julien Tailleur for advice about the implementation of the numerical
algorithm. 
A.~R.\ acknowledges financial support from the Hungarian Scientific Research Fund (Grant No.\ OTKA T-043734).


\appendix
\def\thesection{\Alph{section}}
\section*{Appendix}

\section{Analytical calculations for single site model}
\label{appendix}
\subsection{Spectrum}

\label{ss:spectrum} Naively, one expects that the large deviation
of current fluctuations is given by the lowest eigenvalue of $\tilde{H}$
according to (\ref{eq:Htilde}). For this reason, we here study the
spectrum of (\ref{eq:tildeH}). First we transform $\tilde{H}$ into
the symmetric form 
\begin{equation}
\tilde{H}'(\lambda)=\Phi H(\lambda)\Phi^{-1}
\end{equation}
using the diagonal operator
\begin{equation}
\Phi=\left(\begin{array}{cccc}
\phi & 0 & 0 & \cdots\\
0 & \phi^{2} & 0 & \cdots\\
0 & 0 & \phi^{3} & \cdots\\
\vdots & \vdots & \vdots & \ddots.\end{array}\right),\quad\textrm{with }\phi=\sqrt{\frac{\gamma \rme^{\lambda}+\beta}{\alpha \rme^{-\lambda}+\delta}}.
\end{equation}
 The diagonalization of $\tilde{H}'(\lambda)$ is easily done by a
Fourier transformation. We introduce $y(\lambda)=\sqrt{\left(\gamma \rme^{\lambda}+\beta\right)\left(\alpha \rme^{-\lambda}+\delta\right)}/\left(\beta+\gamma\right)$
which determines the character of the spectrum as follows:

\begin{lyxlist}{00.00.0000}
\item [{$y(\lambda)>1$}] Here the spectrum is entirely continuous. The
eigenvector $|\psi'(k)\rangle$ corresponding to wave number $k\in(0,\pi]$
takes the form
\begin{equation} \fl
|\psi'(k)\rangle=\sqrt{\frac{2}{\pi}}\sum_{n=0}^{\infty}\sin(kn+\varphi)|n\rangle,\quad\textrm{with }\rme^{2i\varphi}=\frac{y(\lambda)\rme^{ik}-1}{y(\lambda)\rme^{-ik}-1},\label{eq:varphi}
\end{equation}
and the corresponding eigenvalue is 
\begin{equation} \fl
\epsilon(k)=\alpha+\beta+\gamma+\delta-2\sqrt{\left(\gamma \rme^{\lambda}+\beta\right)\left(\alpha \rme^{-\lambda}+\delta\right)}\cos k.
\end{equation}

\item [{$y(\lambda)<1$}] In addition to the above continuous spectrum
a discrete band appears with 
\begin{equation} \fl
|\psi'(0)\rangle=\sqrt{1-y^{2}}\sum_{n=0}^{\infty}y^{n}|n\rangle\quad\textrm{and }\epsilon(0)=\alpha+\delta-\left(\beta+\gamma\right)y^{2}.
\end{equation}
 Here this is the lowest eigenvalue, correspondingly the spectrum
becomes gapped. Notice that $\epsilon({k\to 0})\neq\epsilon(0)$.
\end{lyxlist}
We note that it is easy to prove that the above set is complete, i.e.,
\begin{equation}
\delta_{m,n}=\left\lbrace \begin{array}{ll}
\int_{0}^{\pi}\langle m|\psi'(k)\rangle\langle\psi'(k)|n\rangle \,
\rmd k & y(\lambda)>1 \\
\int_{0}^{\pi}\langle m|\psi'(k)\rangle\langle\psi'(k)|n\rangle \,
\rmd k+\langle m|\psi'(0)\rangle\langle\psi'(0)|n\rangle & y(\lambda)<1 
\end{array}\right.  \label{eq:complete}
\end{equation}

The right and left eigenvectors $|\psi(k)\rangle$ and $\langle\psi(k)|$
of $\tilde{H}(\lambda)$ can be obtained by applying the operator
$\Phi$:
\begin{equation}
|\psi(k)\rangle=\Phi^{-1}|\psi'(k)\rangle,\quad\langle\psi(k)|=\langle\psi'(k)|\Phi.\label{eq:transform}
\end{equation}
for $k\in\left[0,\pi\right]$ . In summary, for the lowest eigenvalue $\epsilon_{0}(\lambda)$ (infimum
of the spectrum) of $\tilde{H}(\lambda)$ we obtain
\begin{equation}
\epsilon_{0}(\lambda) = \left\lbrace \begin{array}{ll}
\alpha+\delta-\left(\beta+\gamma\right)y(\lambda)^{2} & y(\lambda)>1\\
\alpha+\beta+\gamma+\delta-2\left(\beta+\gamma\right)y(\lambda) & y(\lambda)<1
\end{array}\right. .
\end{equation}
It can easily be seen that this expression satisfies the symmetry
relation (\ref{eq:esymmetry}), since $y(\lambda)=y(E-\lambda)$ with
$\rme^{E}=\left(\alpha\beta\right)/\left(\gamma\delta\right)$.

\subsection{Calculation of the large deviation function}
\label{ss:ld}

As a first step we write 
\begin{equation}
\langle s|\rme^{-\tilde{H}t}|P_{0}\rangle=(1-x)\sum_{m}\sum_{n}x^{n}\langle m|\rme^{-\tilde{H}t}|n\rangle,
\end{equation}
 where $\langle m|$ is a row vector with a `1' at the $m$th position
and `0's elsewhere. Inserting a complete set of eigenvectors we can
write this in the form 
\begin{eqnarray}\fl
\langle
s|\rme^{-\tilde{H}t}|P_{0}\rangle=(1-x)\sum_{m}\sum_{n}x^{n}\int_{0}^{\pi}\langle
m|\psi(k)\rangle\langle\psi(k)|n\rangle \rme^{-\epsilon(k)t} \, \rmd k\cr
+\theta\left(1-y\right)(1-x)\sum_{m}\sum_{n}x^{n}\langle m|\psi(0)\rangle\langle\psi(0)|n\rangle \rme^{-\epsilon(0)t}
\end{eqnarray}
 Here $\theta$ denotes the Heaviside function. Using the $k\to-k$
symmetry of the eigenvectors and eigenvalues in the contribution of the continuous part, the above integral can be rewritten in the form 
\begin{equation} \fl
\int_{0}^{\pi}\langle m|\psi(k)\rangle\langle\psi(k)|n\rangle
\rme^{-\epsilon(k)t}\, \rmd k=\frac{1}{2\pi}\phi^{n-m}\int_{0}^{2\pi}\left(\rme^{\rmi
  k(n-m)}-\rme^{2\rmi\varphi}\rme^{\rmi
  k(n+m)}\right)\rme^{-\epsilon(k)t}\, \rmd k.
\end{equation}
 After the substitution $z=\rme^{\rmi k}$ and using (\ref{eq:varphi}) we
obtain 
\begin{eqnarray} \fl
\langle
s|\rme^{-\tilde{H}t}|P_{0}\rangle=(1-x)\sum_{m}\sum_{n}x^{n}\phi^{n-m}\frac{1}{2\pi
  \rmi}\oint_{\left|z\right|=1}\left(z^{n-m-1}+\frac{yz-1}{z-y}z^{n+m}\right)\rme^{-\varepsilon(z)t}\, \rmd z\cr
+\theta\left(1-y\right)(1-x)\sum_{m}\sum_{n}x^{n}\phi^{n-m}\left(1-y^{2}\right)y^{n+m}\rme^{-\epsilon(0)t}, \label{eq:poles}
\end{eqnarray}
where by $\varepsilon(z)$ we mean $\epsilon(k(z))$ based on the above substitution.
In order to be able to perform the infinite sums in (\ref{eq:poles}),
we have to choose the contour of the integral carefully. In the first
term of the integral there is a pole only at $z=0$. Here we chose
the contour to be a circle of radius $\phi^{-1}<\left|z\right|<\left(\phi x\right)^{-1}$ (denoted
by $C_{1}$). In the second term we deform the contour to run along a circle around the origin with infinitesimal radius (denoted by $C_{2}$). Notice that doing
so, for $y<1$ we pick up a pole contribution at $z=y$, which just
cancels with the last term in (\ref{eq:poles}) (since
$\epsilon({0})=\varepsilon(y)$).
Finally we obtain 
\begin{eqnarray} \fl
\langle s|\rme^{-\tilde{H}t}|P_{0}\rangle=-\frac{1-x}{2\pi \rmi
  x\phi}\oint_{C_{1}}\frac{\rme^{-\varepsilon(z)t}}{\left(z-\frac{1}{x\phi}\right)\left(z-\frac{1}{\phi}\right)}
  \, \rmd z \cr
+\frac{1-x}{2\pi \rmi
  x}\oint_{C_{2}}\frac{\left(yz-1\right)\rme^{-\varepsilon(z)t}}{\left(z-\frac{1}{x\phi}\right)\left(z-\phi\right)\left(z-y\right)}
  \, \rmd z.\label{eq:final}
\end{eqnarray}
To obtain the large deviation function we need to study the limit of
this integral as $t \to \infty$.  This enables us to build up the
phase diagram as shown in section~\ref{s:phase_diagrams}.

\section*{References}
\bibliographystyle{myunsrtb}
\bibliography{myrefs,allref}

\end{document}